\documentclass[authoryear]{elsarticle}
\usepackage{times}
\usepackage[authoryear]{natbib}
\usepackage{float}
\usepackage[]{graphicx}
\usepackage[dvipsnames]{xcolor}
\usepackage{amsthm}
\usepackage{array}
\usepackage{url}		
\usepackage{xspace}		
\usepackage{booktabs}
\usepackage{ulem}
\usepackage{rotating}	
\usepackage{pdflscape}  
\usepackage{fancyvrb}   
\usepackage{epstopdf}
\usepackage{amssymb}
\usepackage{multirow}
\usepackage{hhline}

\usepackage{enumitem}

\usepackage{optidef} 

\usepackage[pdftex,                %
bookmarks         = true,
bookmarksnumbered = true,
pdfpagemode       = None,
pdfstartview      = FitH,
pdfpagelayout     = SinglePage,
colorlinks        = true,
linkcolor		  = magenta,
citecolor		  = blue,
urlcolor          = orange,
pdfborder         = {0 0 0}
]{hyperref}

\makeatletter
\newcommand{\addresseshere}{%
	\enddoc@text\let\enddoc@text\relax
}
\makeatother
\chardef\bslash=`\\ 

\renewcommand{\emph}[1]{{\it #1}}
\newcommand{\mbf}{\mathbf}
\newcommand{\bias}{\textnormal{Bias}}
\newcommand{\var}{\textnormal{Var}}
\newcommand{\mse}{\textnormal{MSE}}
\newcommand{\E}{\textnormal{E}}

\newcommand{\R}{\mathbb{R}}
\newcommand{\N}{\mathbb{N}}
\renewcommand{\l}{\ell}
\renewcommand{\P}{\mathbf{P}}
\newcommand{\cH}{\mathcal{H}}

\newcommand{\Bin}{\mathbf{Bin}}

\newcommand{\ind}[1]{{\mbf{1}\{#1\}}}
\newcommand{\unifrv}{\mathcal{U}}

\newcommand{\FDR}{\textnormal{FDR}}

\newcommand{\mPCNew}{\widehat{m}_0^{\text {PC,new}}}

\newcommand{\mPCOrig}{\widehat{m}_0^{\text {PC,2006}}}

\newcommand{\mStorey}{\widehat{m}_0^{\text {Storey}}}

\newcommand{\mrand}{\widehat{m}_0^{\text {rand}}}
\newcommand{\mrandMC}{\widehat{m}_0^{\text {rand-MC}}}
\newcommand{\NMC}{N_{\text {MC}}}

\newcommand{\mZZKB}{\widehat{m}_0^{\text {PC,ZZD}}}

\newcommand{\mresc}{\widehat{m}_0^{\text {resc}}}
\newcommand{\mmean}{\widehat{m}_0^{\text {mean}}}

\newcommand{\piZeroStorey}{\widehat{\pi}_0^{\text {Storey}}}

\newcommand{\piZerorandMC}{\widehat{\pi}_0^{\text {rand-MC}}}

\newcommand{\piZeroConv}{\hat{\pi}_0^{{\mathcal G} (\blambda)}}
\newcommand{\gNewStar}{g^{\text {new},\ast}}

\newcommand{\gStoreyStar}{g^{\text {Storey},\ast}}

\newcommand{\blambda}{\boldsymbol{\lambda}}
\newcommand{\bw}{\boldsymbol{w}}

\newcommand{\gest}{\ge_{\text{st}}}
\newcommand{\lest}{\le_{\text{st}}}
\newcommand{\lecx}{\le_{\text{cx}}}
\newcommand{\gecx}{\ge_{\text{cx}}}
\newcommand{\cxorder}{\lecx}
\newcommand{\stoorder}{\lest}

\newcommand{\nullset}{\mathcal{H}_{0}}
\newcommand{\altset}{\mathcal{H}_{1}}

\graphicspath{{img/}}

\newtheorem{theorem}{Theorem}[section]

\newtheorem{proposition}{Proposition}[section]

\newtheorem{definition}{Definition}[section]
\newtheorem{remark}{Remark}[section]
\newtheorem{example}{Example}[section]
\newtheorem{lemma}{Lemma}[section]

\begin{document}
\title{ Improved null proportion estimators for multiple discrete  tests with plug-in FDR control 
}
\tnotetext[t1]{This document is the result of a research 	project funded by the Deutsche Forschungsgemeinschaft (DFG).}

\author[1]{Iqraa Meah}
\ead{iqraa.meah@agroparistech.fr}

\author[2,3]{Sebastian D\"ohler\corref{cor1}}
\ead{sebastian.doehler@h-da.de}

\cortext[cor1]{Corresponding author}

\affiliation[1]{organisation={UMR MIA Paris-Saclay, AgroParisTech, INRAE, Université Paris-Saclay},
city={Palaiseau},
postcode={91120},
country={France}}

\affiliation[2]{organisation={Darmstadt University of Applied Sciences},
addressline={Schöfferstr. 3},
city={Darmstadt},
postcode={64295},
country={Germany}}
\affiliation[3]{organisation={EUt+ Data Science Institute},
	addressline={European University of Technology},
	country={European Union}}


\begin{abstract}
It is well known that the performance of the Benjamini-Hochberg (BH) procedure can be improved by incorporating estimators of the number or proportion of null hypotheses to yield an adaptive BH procedure which still controls FDR. Several such plug-in estimators have been proposed. For some of these, such as Storey's estimator, plug-in FDR control has been established, while for some others, such as  the Pounds–Cheng estimator, some gaps remain to be closed.
These developments have largely focused on the case of continuous test statistics, where null $p$-values follow the uniform distribution. In the discrete setting, although these estimators continue to provide plug-in FDR control, they  become overly conservative, leading to inefficient procedures.
In this paper, a general class of estimators that encompasses the classical Storey and Pounds–Cheng estimators is introduced. Alongside, several generic strategies to mitigate conservativeness in the discrete setting are proposed by incorporating information about the null distribution functions.
These strategies provably yield less conservative estimates while maintaining valid FDR control, and the resulting performance gains are illustrated on both real and simulated data. As a byproduct of a more general result,  plug-in FDR control for the Pounds–Cheng estimator in the continuous case is also established.

\end{abstract}


\maketitle

\section{Introduction}\label{sec:intro}
\subsection{Background} \label{ssec:background}
When many statistical tests are performed simultaneously, a ubiquitous way to account for the false rejections is the false discovery rate (FDR), that is the expected proportion of errors among the rejections. 
The seminal \cite{BH95} procedure (abbreviated in the sequel as BH procedure) works by rejecting $H_{(1)}, \ldots, H_{(\hat{k})}$, where $\hat{k}$ is determined in the following \emph{step-up} manner
\begin{align}
	\hat{k} &= \max \left\{ \l \in \{0, \ldots, m \} : p_{(\l)} \le \frac{\l}{m} \cdot \alpha \right\}, \label{eq:def:plug:in:BH}
\end{align} 
where $ p_{(1)} \le \ldots \le p_{(m)}$ denote the ordered $p$-values, $H_{(1)} , \ldots, H_{(m)}$ the corresponding null hypotheses and  $p_{(0)} := 0$. 
According to results in \cite{BH95} and \cite{BY2001}, this procedure guarantees that $\FDR \le \pi_{0}\alpha$ when the $p$-values are independent or positively dependent, while
for arbitrarily dependent $p$-values the \cite{BY2001} procedure is available. The simplicity of the BH procedure  and its many useful theoretical properties have made it an indispensable tool in modern high dimensional data analysis, see e.g. \cite{benjamini2010simultaneous}.

Much work has gone into analyzing, extending and adapting this procedure to various settings. For instance, \cite{Storey2004} showed that the \emph{plug-in BH procedure}
\begin{align}
	\hat{k} &= \max \left\{ \l \in \{0, \ldots, m \} : p_{(\l)} \le \frac{\l}{\widehat{m}_0} \cdot \alpha \right\}, \label{eq:khat:BH} 
\end{align} 
obtained by replacing  $m$ in \eqref{eq:def:plug:in:BH} by an estimate $\widehat{m}_0$ of $m_0$ still provides so-called \emph{adaptive} or \emph{plug-in} $\FDR$ control at level $\alpha$ while often allowing for more power. 
Classical examples of such estimates were proposed by \cite{Storey2004}
\begin{align}
	\mStorey&= \frac{1 + \sum_{i =1}^m \ind{p_i > \lambda}}{1-\lambda}, \label{def:m0:Storey}\\
	\intertext{where $\lambda \in [0, 1)$ is a tuning parameter, and by \cite{PC2006} }
	\mPCOrig&= m \wedge  \left( 2\sum_{i =1}^m  p_i \right). \label{def:m0:PC:2006} 
\end{align}

While the focus of this work is on plug-in FDR control, estimates of $m_0$ (or equivalently $\pi_0=m_0/m$) can also be used for FDR estimation purposes (see \cite{Storey2002}). 
Thus, there is a large body of literature on this topic and numerous methods for establishing plug-in FDR control are available, see e.g. \citet{benjamini2006adaptive, Sarkar2008, BR2009, heesen2016dynamic, DitzhausJanssen} and references therein. 
{In this paper, we use a classical condition proposed by \cite{BR2009} for establishing plug-in FDR control for a unified class of $m_0$-estimators, see Section~\ref{ssec:imc} for more details.}
{Previous work on plug-in FDR control has focused on} continuous test statistics, for which the null $p$-values are distributed according to the uniform distribution. 
{Considering the abundance of super-uniform $p$-values in real-life applications, the uniformity assumption is often violated which may lead to undesirable conservatism of the $m_0$-estimators, see Section~\ref{sec:discretestimators} for more details.}
{Super-uniform $p$-values can be observed when testing composite null hypotheses or when dealing with discrete tests, the latter being the focus of this work.}

Discrete tests often originate when the tests are based on counts or contingency tables:
for example in clinical studies, the efficacy or safety of drugs is determined by counting patients who survive a certain period, or experience a certain type of adverse drug reaction after being treated, see e.g. \cite{chavant2011memory}; 
and also in biology, where the genotype effect on the phenotype can be analyzed by knocking out genes and counting the number of individuals with a changed phenotype, see e.g. \cite{munoz2018international}.
In discrete testing, each $p$-value is super-uniform and (potentially) has its own support, thus producing heterogeneous $p$-values. 
{The analysis of the latter represents an important part of contemporary statistical research, see e.g. \cite{OchiengHoangDickhaus2024} and the references cited therein.
In the context of $\FDR$ estimation,} \cite{PC2006} recognized the need for developing methods tailored  to discrete $p$-values and  introduced $\mPCOrig$  as a simple and robust $m_0$-estimate in this setting. 
They did not, however, provide a proof of plug-in FDR control, not even in  the uniform setting. 
Further works addressing the discreteness and heterogeneity include \cite{chen2018multiple} who introduced an $m_0$-estimator {which actively adjusts} for discrete $p$-values based on averaging Storey type estimators for plug-in control. 
{After \cite{BiswasLetter2020} pointed out an error in the proof of \cite{chen2018multiple}, \cite{ChenDoerge2020} introduced an additional condition in an erratum under which plug-in FDR control for this estimator could be guaranteed. This additional constraint may however be quite restrictive in applications. \cite{BiswasChattopadhyay2024} rectified the line of thinking in \cite{chen2018multiple}, concluding that further research is needed to find robust and valid estimators in the discrete paradigm.
Thus, there are gaps to be filled on $m_0$-estimation both for the uniform and discrete case.}

\subsection{Contributions}
{In this paper, we address some of the limitations that occur in the discrete settings by introducing a simple and flexible class of $m_{0}$-estimators. 
More precisely, }
	\begin{itemize}
	
		\item  We introduce three simple approaches to adapt classical estimators to the discrete paradigm where loss of power is encountered due to the over conservativeness of the $p$-values.
			 We investigate the performance of these discrete estimators for several real data sets and in simulations with comparison to their non adapted versions to discreteness.
			
		\item For each of these approaches, we guarantee the plug-in FDR control by introducing a general class of conservative estimators. 
			This class provides a simple and flexible generic formulation {which is useful for designing} new conservative estimators regardless of the $p$-value distribution under independence.
			The proof of plug-in FDR control for this generic estimator is based on simple convex ordering arguments {(see  ~\ref{appendix:auxres})}.


	\end{itemize}

On another note, our generic class of estimators also allows us to show that a simple modification of $\mPCOrig$ is contained in our general class which allows us to obtain a version of $\mPCOrig$ with guaranteed plug-in FDR control closing the gap on this estimator.

The paper is organized as follows: in the remaining of the current section, we present the statistical setting and {restate} {a classical sufficient} criterion for plug-in FDR control. 
Then, Section \ref{sec:uniform} introduces the new class of estimators, and presents the main mathematical results on plug-in FDR control. 
In Section~\ref{sec:discretestimators} we present approaches for adjusting estimators to discreteness, {and investigate their performance on simulated and real data in Sections~~\ref{ssec:simu:discrete} and ~\ref{sec:real:data:analysis}. Section~\ref{sec:outlook} sketches some approaches useful for choosing suitable estimators from our new class and Section~\ref{sec:conclusion} concludes the paper.
{Technical details including classical results on stochastic and convex ordering, {mathematical proofs} and further analyses are deferred to the Appendices}.

\subsection{Distributional assumptions} \label{sec:distribassumption}
We use a classical setting for multiple testing encompassing homogeneous and heterogeneous nulls, see e.g. \cite{DDR2018}. 
We observe $X$, defined on an abstract probabilistic space, valued in an observation space
$(\mathcal{X},\mathfrak{X})$ and generated by a distribution $P$ that  belongs to a set $\mathcal{P}$ of possible distributions. 
We consider $m$ null hypotheses for $P$, denoted $H_{0,i}$, $1 \leq i \leq m$,  and we denote the corresponding set of true null hypotheses by $\cH_0(P)=\{1\leq i\leq m\::\: \mbox{$H_{0,i}$ is satisfied by $P$}\}$. 
We also denote by $\cH_1(P)$ the complement of $\cH_0(P)$ in $\{1,\dots,m\}$ and by $m_0(P) =  m_{0} =|\cH_0(P)|$ the number of true nulls.\\
We assume that there exists a set of $p$-values that is a set of random variables (r.v) $\{p_i(X), 1\leq i \leq m\}$, valued in $[0,1]$. 
We introduce the following dependence assumptions between the $p$-values:}

{
\begin{align}
	\mbox{All the  $p$-values  $\{p_i(X)$, $1\leq i \leq m\}$ are mutually independent in the model $\mathcal{P}$.}
	\label{Indep}\tag{Indep}
\end{align}
}

The (maximum) null cumulative distribution function (c.d.f) of each $p$-value is denoted 
\begin{equation*}
	F_{i}(t) = \sup_{P\in \mathcal{P}\::\: i\in \cH_0(P)} \{\P_{X\sim P}(p_{i}(X)\leq t )\}, \:\: t\in[0,1], \:\:1\leq i\leq m.
\end{equation*}
We assume that the c.d.f.'s $F_1, \ldots, F_m$ are {\it known} and we consider the following possible situations:
\begin{align}
	&\mbox{For all $i \in\{1,\dots,m\}$, $F_i$ is continuous on $[0,1]$.} \tag{Cont} \label{cont}\\
	&\begin{array}{c}	\mbox{For all $i \in\{1,\dots,m\}$, there exists some finite 
	support  $\mathcal{S}_i\subset [0,1]$ such that}\\
					\mbox{$F_i$ is a step function, right continuous, that jumps only at some points of $\mathcal{S}_i$.}
	\end{array}
	\tag{Discrete} \label{discrete}
\end{align}
The case \eqref{discrete} typically arises when for all $P\in \mathcal{P}$ and $i\in\{1,\dots,m\}$, $\P_{X\sim P}(p_i(X)\in \mathcal{S}_i)=1$ {for some given supports $\mathcal{S}_i\subset [0,1]$}.
Throughout the paper, we will assume that we are either in the case \eqref{cont} or \eqref{discrete} and we denote $\mathcal{S}=\cup_{i=1}^m \mathcal{S}_i$, with by convention $\mathcal{S}_i=[0,1]$ when \eqref{cont} holds.
We will also make use of the following classical assumption:
\begin{align}
	&\mbox{For all $i \in\{1,\dots,m\}$,  $F_{i}(t)\leq t$ for all $t\in[0,1]$}.\tag{SuperUnif} \label{superunif}
\end{align}
In this paper, we will always assume that the $p$-values are mutually independent (\eqref{Indep} holds) and super-uniform under the null (\eqref{superunif} holds). 

\subsection{FDR control for plug-in estimates} \label{ssec:imc}
The following Theorem is a central result for plug-in FDR control, providing a sufficient condition based on bounding the inverse moment of the estimator $\widehat{m}_0$ by the inverse of $m_0$.  
Our presentation follows \cite{BR2009}, similar results can be found in \cite{benjamini2006adaptive, Sarkar2008, ZZD2011}.

\begin{theorem} \label{thm:IMC}
Let $\widehat{m}_0 = \widehat{m}_0(p_1, \ldots, p_m)$ be a coordinatewise non-decreasing function of the $p$-values $(p_1, \ldots, p_m)$. 
	Assume that $(p_1, \ldots, p_m)$ are mutually independent \eqref{Indep} and \eqref{superunif} holds true. 
	For $h \in \nullset$, denote by $ p_{0, h}$ the set of $p$-values where $p_h$ has been replaced by $0$. If 
	\begin{align}
		\E \left( \frac{1}{\widehat{m}_0 ( p_{0, h})}\right) & \le 	
		\frac{1}{m_0} \label{eq:IMC} \tag{IMC}
	\end{align} 
	holds for all $h \in \nullset$, then the plug-in BH procedure given by \eqref{eq:khat:BH} controls FDR  at level $\alpha$.
\end{theorem}

Throughout this paper, we will only consider coordinatewise non-decreasing estimators and  will always assume that the $p$-values are mutually independent. Thus, when additionally \eqref{superunif} holds true, the inverse moment criterion \eqref{eq:IMC} is {sufficient} for establishing plug-in FDR control in our proofs. 
{We mostly present results {in terms of the}  absolute number of null hypotheses $m_0$, but {clearly} equivalent statements using the proportion of null hypotheses $\pi_0=m_0/m$ hold, and in some cases we will present results in terms of $\pi_0$ instead of $m_0$.}
\section{A general result for plug-in estimators} \label{sec:uniform}

{In this section we present a result on plug-in FDR control for a general class of estimators, which is applicable both to continuous and discrete $p$-values.}
This class of estimators  is based on sums of $p$-values  that have been transformed by bounded non-decreasing functions, which allows us  to recover classical estimators, such as the Storey \eqref{def:m0:Storey} and the PC (slightly modified) \eqref{def:m0:PC:2006} estimators, and also to define new estimators. 

{Let $\unifrv[0,1]$ denote the uniform distribution on $[0,1]$.} To start, assume that the $p$-values are transformed by certain functions $g \in \mathcal{G}$, with 
\begin{align*} 
	\mathcal{G}&= \{g:[0,1] \rightarrow [0,1] : \text{ $g$ is non-decreasing and $\E [g(U)] > 0 $,  where $U \sim \unifrv[0,1]$}\}.
\end{align*} 

Accordingly, we define the class of estimators $\mathcal{F}_0$ as

\begin{align}
	\mathcal{F}_0&= \left\{ \widehat{m}_0 : [0,1]^m \rightarrow  [0,\infty) \vert\: \widehat{m}_0 (p_1, \ldots, p_m) = \frac{1}{{\nu(g)}} \left(1+ \sum_{i=1}^m g(p_i) \right), g \in \mathcal{G}
	\right\}, \label{eq:def:class:estimators:0}
\end{align} 
where $\nu(g)= \E [g(U)]$ for any $g \in \mathcal{G}$ with $U \sim \unifrv[0,1]$ (for brevity we sometimes omit the $g$ in $\nu$ when there is no ambiguity concerning the function $g$).
The class $\mathcal{F}_0$ contains the classical estimator $\mStorey$ \eqref{def:m0:Storey} by taking $g(u)=\ind{u>\lambda}$ and  $\nu=1-\lambda$. 
It also contains a slightly modified version $\mPCNew$ of the classical estimator $\mPCOrig$ \eqref{def:m0:PC:2006} obtained from taking $g(u)=u$ with $\nu=1/2$, i.e. 
\begin{align}
	\mPCNew &= 2 + 2 \sum_{i=1}^{m} p_i  \label{eq:def:PC:new}.
\end{align}
{See ~\ref{appendix:ssec:ComparePCNew:PCZZD} for more details on $\mPCNew$.}
{The rationale behind the definitions of the classes $\mathcal{G}$ and $\mathcal{F}_0$ is two-fold. Requiring that $g$ is non-decreasing ensures that $\widehat{m}_0$ is coordinatewise non-decreasing, allowing  us to apply Theorem~\ref{thm:IMC}. 
The quantity $g(p_i)/\nu$ can be interpreted as the (local) contribution of $p_i$ to the estimate of $m_0$. 
If we expect large $p$-values to provide evidence for null hypotheses, then it seems reasonable to require $g$ to be non-decreasing. Rescaling $g(p_i)$ by $\nu= \E [g(U)]$ is a simple way of ensuring that $\sum_{i=1}^m g(p_i)/\nu$ is conservatively biased in the sense that $\E(\sum_{i=1}^m g(p_i)/\nu)\ge m_0$ in any constellation of null and alternative hypotheses. 
This type of conservativeness may however not be strong enough for plug-in control. 
As our main result -- Proposition~\ref{prop:plugin:control:general:g} below -- shows, simply adding $1/\nu$ as a 'safety margin' to the above estimate is enough for ensuring plug-in FDR control.}\\
{In the analysis of multiple discrete tests, the $p$-values under the null may be heterogeneous, i.e. the $p$-values may  have different distributions under the null, so that using an individual transformation for each $p$-value will prove to be useful. 
To this end, we introduce the following  richer and more flexible class of estimators. }
\begin{equation}
	\begin{split}
		\mathcal{F} = \biggl\{ \widehat{m}_0 : [0,1]^m &\rightarrow [0,\infty) \ \bigg\vert \ \widehat{m}_0 (p_1, \ldots, p_m) = \frac{1}{\min(\nu_1, \ldots, \nu_m)} \\
		&+ \sum_{i=1}^m \frac{g_i(p_i)}{\nu_i}, \quad \text{with $g_i \in \mathcal{G}$ and $\nu_{i} = \E[g_{i}(U)]$, $U \sim \unifrv[0,1]$ for all $i$} \biggr\}.
	\end{split}
	\label{eq:def:class:estimators}
\end{equation}
{Clearly,  $\mathcal{F}_0 \subset \mathcal{F}$, so that any result on plug-in FDR control that holds for $\mathcal{F}$ {also holds} for $\mathcal{F}_0$.} 
We therefore state our main result for this {more general} class {in the following Proposition.} 

\begin{proposition}\label{prop:plugin:control:general:g}	
	Assume that $p_1, \ldots, p_m$ are mutually independent and \eqref{superunif} holds. 
	Then \eqref{eq:IMC} holds true for any estimator $\widehat{m}_0 \in \mathcal{F}$, where $\mathcal{F}$ is defined by \eqref{eq:def:class:estimators}. 
	In particular, the BH plug-in procedure \eqref{eq:khat:BH} using $\widehat{m}_0$ controls FDR at level $\alpha$.
\end{proposition} 


In particular,  Proposition~\ref{prop:plugin:control:general:g} immediately implies  plug-in FDR control  of $\mPCNew$ in the classical paradigm of uniform $p$-values, which is a new result (see \ref{appendix:ssec:ComparePCNew:PCZZD} for more details). Our main interest in this paper is, however, to construct improved  estimators for discrete $p$-values. We will use  Proposition~\ref{prop:plugin:control:general:g} to this end  in the next Section. 

\section{Adjusted estimators for discrete $p$-values} \label{sec:discretestimators}

\begin{center}
	\begin{figure}[h!]  
		\center
		\includegraphics[width=.7\linewidth]{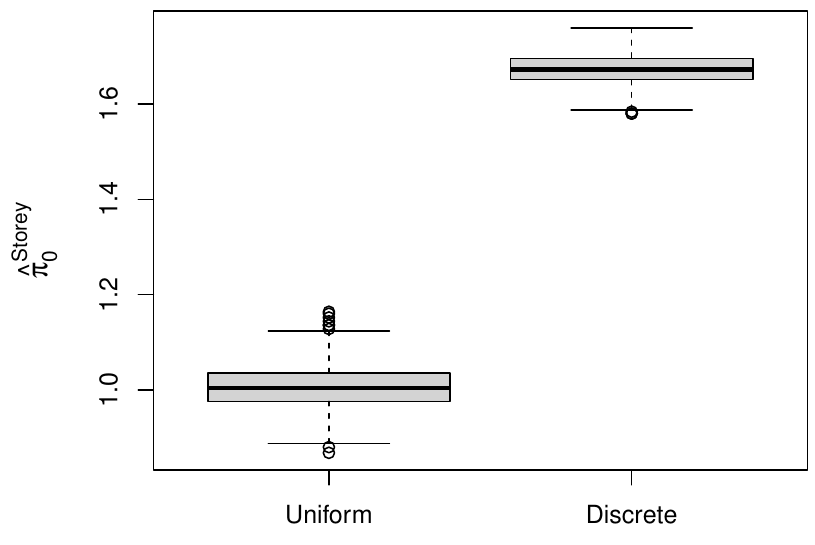}
		\caption{Boxplots of  $\mStorey / m$ on simulated independent data for $m=m_0=500$ over 1000 simulations of uniform and discrete $p$-values. Uniform $p$-values are generated from the $\unifrv[0,1]$ distribution and the discrete $p$-values are generated as described in Section~\ref{ssec:simu:discrete} with $N=25$ and  $\Bin(1, 0.10)$ at $m_3$ positions for both groups to simulate a complete null setting. }
		\label{fig:Storey_unif_vs_discrete}
	\end{figure}
\end{center}

For any estimator $\widehat{m}_0 \in \mathcal{F}_0$ as in \eqref{eq:def:class:estimators:0} the bias is given by 
\begin{align}
	\bias (\widehat{m}_0) &= \E \widehat{m}_0 - m_0 = \frac{1}{\nu(g)} + \sum_{i \in \nullset } \left(\frac{\E g(p_i)}{\nu(g)}-1\right) + \sum_{i \in \altset } \frac{\E g(p_i)}{\nu(g)}. \label{eq:bias:m0:general}
\end{align}
Classical plug-in estimators like $\mStorey$ and $\mPCNew$ were developed for uniformly distributed $p$-values under the nulls so that $\E g(p_i)/\nu(g)=1 $ for $i \in \nullset$ and  therefore the contribution of $\nullset$ in \eqref{eq:bias:m0:general} disappears. In the discrete case however, the $p$-values are super-uniform under  $\nullset$, so that $U \lest p_i$ for $U \sim \unifrv[0,1]$ and therefore $g(U)\lest g(p_i)$, since $g$ is non-decreasing. Thus, for discrete data we have in general $\E g(p_i)/\nu(g)\ge 1 $ for $i \in \nullset$, causing a bias-inflation  of $\widehat{m}_0$ in \eqref{eq:bias:m0:general}. Figure~\ref{fig:Storey_unif_vs_discrete}, presenting estimation boxplot of $\mStorey / m$ under the complete null setting ($m_{0} = m$, i.e. $\pi_{0} = 1$), illustrates this behavior very well. The median of the boxplot in the uniform case is correctly located at 1 while in the discrete case it reaches 1.6. This gap  quantifies the 'null-bias' in \eqref{eq:bias:m0:general} caused by discreteness. Details on the data generating mechanism are given in the figure caption.

In this section we  incorporate the discrete null distributions $F_1, \ldots, F_m$ into three different modifications of  any estimator $\widehat{m}_0\in \mathcal{F}_0$ which improve efficiency in the discrete setting. We will show that these modified estimators are always better than $\widehat{m}_0$  without compromising FDR control.

For a discrete super-uniform $p$-value $p \sim F$  on $[0,1]$ let the support be given by $\mathcal{S}=\{0 <s_1 < s_2 < \ldots < s_L=1\}$ (and for convenience we set $s_0=0$). In what follows we assume throughout that the assumptions \eqref{Indep}, \eqref{discrete} with supports given by $\mathcal{S}_1, \ldots, \mathcal{S}_m$  and \eqref{superunif} are true and the null distribution functions $F_1, \ldots, F_m$ are known.

\subsection{Adjusting the rescaling constants}\label{ssec:adjusting:rescaling}
The idea of this approach is to replace the global normalizing constant $\nu(g)$ by individual normalizing constants $\nu_i$ ensuring that $\E g(p_i)/\nu_i=1 $ for $i \in \nullset$ so that the contribution of nulls to the bias in \eqref{eq:bias:m0:general} disappears. A similar rationale  is described briefly but without further elaboration for the PC estimator in the Discussion Section of \cite{PC2006} in the context of FDR estimation. Here we present a   result for plug-in procedures in the general class of estimators \eqref{eq:def:class:estimators:0}.


\begin{proposition}\label{prop:plugin:discrete:control:general:g}	
	For any $\widehat{m}_0 \in \mathcal{F}_0$ as in \eqref{eq:def:class:estimators:0} with $g \in \mathcal{G}$ define the \emph{rescaled} estimator
	\begin{align}
		\mresc (p_1, \ldots, p_m)&= \frac{1}{min(\nu^{\text{resc}}_1, \ldots, \nu^{\text{resc}}_m ) }+ \sum_{i=1}^m \frac{g(p_i)}{\nu^{\text{resc}}_i}, \label{eq:def:discrete:rescaled:estimator}
	\end{align}
	where $\nu^{\text{resc}}_i = \E_{p_i \sim F_i} [g(p_i)]$, is the expectation {of the transformed p-value}  taken w.r.t. $F_i$. 
	Then the BH plug-in procedure \eqref{eq:khat:BH} using $\mresc $ controls  FDR at level $\alpha$ and $\mresc \le \widehat{m}_0$ (a.s.).
\end{proposition}

For classical estimators, the rescaling constants can be computed easily, i.e. for Storey's estimator  with $g(u)=\ind{u>\lambda}$ we have $\nu^{\text{resc}}_i=1-F_i(\lambda)$ and for the Pounds-Cheng  estimator  with $g(u)=u$ we obtain $\nu^{\text{resc}}_i=\sum_{s \in \mathcal{S}_i} s \cdot P (p_i=s)$.

\subsection{Adjusting the $p$-value transformations by  (conditional) mean values} \label{ssec:mean:value}
The idea of this approach is to replace the global transformation function  $g\in \mathcal{G}$  by some  function $\bar{g}_{F}$ depending on $F$, so that the `deflated' r.v. $\bar{g}_{F}(p)\le g(p)$ mimics more closely the behavior of $g(U)$ for $U \sim \unifrv [0,1]$ in the sense that 
\begin{align}
	\E \bar{g}_{F}(p) &= \E g(U) \label{eq:calibration:g:bar}.
\end{align}
By `deflating' $g(p)$ to $\bar{g}_{F}(p)$,  the normalising constant $\nu(g)$ remains  unchanged, so that  the contribution of nulls to the bias in \eqref{eq:bias:m0:general} disappears.  If $g(u)=u$ then a well-known solution to \eqref{eq:calibration:g:bar} is given by 
\begin{align}
	\bar{g}_{F}(s_k)&=\frac{1}{2} (s_k+s_{k-1})=s_k - \frac{1}{2} (s_k-s_{k-1}), \label{eq:def:midp}
\end{align}
which is the  so-called \emph{mid-$p$-value} associated with the original $p$-value at value $s_k$, see \cite{berry1995mid} for more details, and \citet{chen2020false, chen2020benjamini} for some recent works on FDR control using mid-$p$-values. Traditionally, mid $p$-values have been a popular way of dealing with the conservatism of individual discrete $p$-values, their distribution however, is no longer super-uniform but shrunk toward 0.
For more general functions $g\in \mathcal{G}$ and super-uniform distributions $F$ we ensure that condition \eqref{eq:calibration:g:bar} is satisfied by defining  
\begin{align}
	\bar{g}_{F}(t)&= \sum_{k=1}^{L} a_k \cdot \ind{(s_{k-1},s_k]}(t) \qquad 	\text{with} \qquad 	a_k = \frac{1}{F(s_k)-F(s_{k-1})} \int_{F(s_{k-1})}^{F(s_k)} g(s) ds. \label{def:g:bar:F}
\end{align}
Thus, the random variable $\bar{g}_{F}(p)$ can be seen as a kind of general mid or mean $g(p)$ value. The range of $\bar{g}_{F}$ is given by $a_1,\ldots,a_L$ (and zero), where each $a_k$ is the conditional expectation of $g(U)$ conditioned on  $U \in (F(s_{k-1}),F(s_k)]$.  Additionally, when assuming that discrete $p$-values are  constructed in the usual way, i.e. for $x$ belonging to the support  we have $F(x)=x$,  this yields the simple representation $a_k = \E g(\unifrv(s_{k-1},s_k))$. 
For Storey's estimator  with $g(u)=\ind{u>\lambda}$ we then obtain
\begin{align*}
	a_k &=
	\begin{cases}
		0 & \lambda > s_k,\\
		\frac{s_k - \lambda}{s_k - s_{k-1}} & \lambda \in (s_{k-1},s_k] , \\
		1 & \lambda \le s_{k-1}.
	\end{cases}
\end{align*}
Thus,  we have $g(s_k)=\bar{g}(s_k)$ except when  $\lambda \in (s_{k-1},s_k]$. For the Pounds-Cheng  estimator  with $g(u)=u$ we obtain
\begin{align*}
	a_k &= \frac{1}{2} (s_k+s_{k-1})=s_k - \frac{1}{2} (s_k-s_{k-1}),
\end{align*}
which is identical with the mid-$p$ value associated with observing $s_k$, see \eqref{eq:def:midp}.

 \begin{proposition}\label{prop:plugin:discrete:control:mean:g}	
 	For any $\widehat{m}_0 \in \mathcal{F}_0$ as in \eqref{eq:def:class:estimators:0} with  $g \in \mathcal{G}$ define the (conditional) \emph{mean} estimator
 	\begin{align}
 		\mmean (p_1, \ldots, p_m)&= \frac{1}{\nu(g)} \left(1+ \sum_{i=1}^m \bar{g}_{F_i}(p_i) \right), \label{eq:def:discrete:adjusted:estimator}
 	\end{align}
 	where each $\bar{g}_{F_i}$ is defined as in \eqref{def:g:bar:F}. 
 	Then the BH plug-in procedure \eqref{eq:khat:BH} using $\mmean$ controls  FDR at level $\alpha$ and $\mmean\le \widehat{m}_0 $ (a.s.).
 \end{proposition}

%
\subsection{An approach based on expected randomization} \label{ssec:randomization:approach}
The mean-value approach described in Section \ref{ssec:mean:value} deals with the conservativeness of discrete $p$-values at the level of the transformation function by individually constructing  $\bar{g}_{F_i}$ so that the random variables  $\bar{g}_{F_i}(p_i)$ and $g(U)$ have the same expectation. Another, more fundamental way of dealing with the conservativeness issue at the level of the individual $p$-value are so-called   \emph{randomized $p$-values} which are defined in our context by
\begin{align}\label{eq:def:randomp}
	r(x,u) &= p(x) - u \cdot P_0 (p(X)=p(x)),
\end{align}
where $u$ is the realization of a uniform random variable $U \sim \unifrv[0,1]$, independent of $X$. 
Alternatively to the notation $r(x,u)$, we will also use (with a slight abuse) $r(p,u)$, where $p=p(x)$ is the standard discrete $p$-value obtained from observation $x$.
Randomized $p$-values and mid-$p$ values are related via the formula $p(x)=\E_U [r(x,U) | X = x]$ (here $p(x)$ denotes the mid-p value).
{Thus, randomization leads to an (unconditional) uniform behavior, however at the cost of an additional source of randomness which makes its use controversial for decisions on individual hypotheses, see e.g. \cite{habiger2011randomised} for a discussion. }
In what follows, we show that for estimation purposes, {randomized $p$-values can nevertheless be beneficial for obtaining an efficient non-randomized estimator.}

For discrete data, \cite{dickhaus2012analyze} argue for using randomized $p$-values in (essentially) Storey's estimator, i.e. applying $\mStorey$ to $(r_1, \ldots,r_m)$ instead of $(p_1, \ldots, p_m)$ which yields a random estimate that should provide  a less conservative estimate for $m_{0}$. 
They show that plugging this estimator  into the Bonferroni procedure yields asymptotic control of the Familywise Error Rate (FWER) under certain assumptions. 
They also point out that if fully reproducible results are desired it may be more appropriate to work with the conditional expectation w.r.t. randomization, i.e. using $\E_{(U_1, \ldots, U_m)} \left(\mStorey (r_1, \ldots,r_m)\right)$ instead of $\mStorey (r_1, \ldots,r_m)$.
We follow and extend this idea by introducing an \emph{expected randomized estimator} which is defined for  any estimator $\widehat{m}_0$, not necessarily belonging to $\mathcal{F}_0$  as 
{
\begin{align}
	\mrand (p_1, \ldots, p_m) &=  \left[ \E_{U} \left(\frac{1}{\widehat{m}_0 (r(p,U))}\right)\right]^{-1} \label{def:m0:rand}
\end{align}
with $U=(U_1, \ldots, U_m )$ where $U_1, \ldots, U_m \sim \unifrv[0,1]$ denote i.i.d. uniform random variables independent of $p= (p_1, \ldots, p_m)$ and we write $\widehat{m}_0(r(p,U))$ instead of $\widehat{m}_0 (r_1(p_1,U_1), \ldots, r_m( p_m,U_m))$ for brevity.} Thus, for fixed $(p_1, \ldots, p_m)$ this estimator is obtained by taking the expectation over the randomized $p$-values associated with $(p_1, \ldots, p_m)$. Again, this approach comes with guaranteed FDR plug-in control.

{
	\begin{proposition} \label{coro:randomized:plugin:control}
		Let $\widehat{m}_0$   {satisfy the conditions of Theorem \ref{thm:IMC}}. 
	\begin{itemize}
			\item[(a)] 	Then the BH plug-in procedure  \eqref{eq:khat:BH} using $\mrand (p_1, \ldots, p_m) $ defined by \eqref{def:m0:rand} controls  FDR at level $\alpha$. 
			\item[(b)] $\mrand \le \widehat{m}_0$ (a.s.).
			\item[(c)] For any $\widehat{m}_0 \in \mathcal{F}_0$ as in \eqref{eq:def:class:estimators:0} with  $g \in \mathcal{G}$ we have $\mrand \le \mmean$ (a.s.).
		\end{itemize}
	\end{proposition}
}

Proposition  \ref{coro:randomized:plugin:control} shows that we can obtain  plug-in FDR control in a finite-sample setting for {any estimator  $\widehat{m}_0$ satisfying  the conditions of Theorem~ \ref{thm:IMC} and in particular for  $\widehat{m}_0 \in \mathcal{F}_0$ by using conditional expectation w.r.t. randomization.}  
{Like the preceding estimators, this estimator is at least as efficient as the standard (uniform) estimator.}

Additionally, statement (c) implies that $\mrand$ is preferable to $\mmean$ if $\widehat{m}_0 \in \mathcal{F}_0$. In practice however, the conclusion is less clear  since the 'ideal' estimator  \eqref{def:m0:rand} can generally not be determined analytically. This means that we will usually need to approximate $\mrand$ numerically e.g. by  the Monte-Carlo approximation
	\begin{align}
		\mrandMC  (p_1, \ldots, p_m)  &= \left[ \frac{1}{\NMC} \sum_{k=1}^{\NMC} \frac{1}{\widehat{m}_0(r(p,U_k))}\right]^{-1},\label{def:m0:rand:MC}
	\end{align}
	where the vector $U$ is simulated $\NMC$  times. The additional randomness introduced by $\mrandMC$ is somewhat dissatisfying since this is precisely what we sought to avoid.  In any case, if we want to use $\mrandMC$ we should aim to  control the variability of the Monte-Carlo estimator by  appropriate choice of $\NMC$. A pragmatic approach may be to explore the variability of the (simulated) distribution of $\mrandMC $. Figure \ref{fig:StoreyMC} gives an example for the amnesia data  (see  \ref{appendix:sec:additional:analyses:real:data} for more information) where $\NMC=100$ was used and $ \piZerorandMC=\mrandMC/m$ was simulated $1000$ times. \begin{center}
		\begin{figure}[th]  
			\center
			\includegraphics[width=0.8\linewidth]{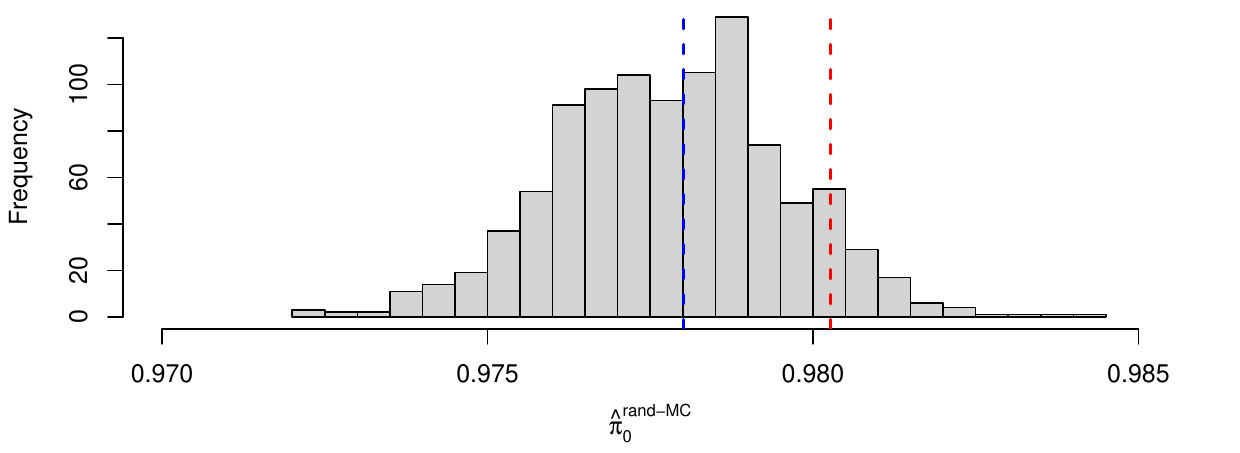}
			\caption{Histogram for 1000 simulations of $\piZerorandMC$ for Storey's estimator ($\lambda=0.5$) with $\NMC=100$ on the amnesia data. The blue vertical line represents  the value of $\mmean$, the red line represents $\mresc$.}
			\label{fig:StoreyMC}
		\end{figure}
	\end{center}
	As the histogram shows, the (true but unknown) value of  $\mrand$ seems to lie in the interval $[0.970,0.985]$. Alternatively, the CLT could be used to determine $\NMC$ in order to obtain a pre-specified precision.

The slightly complicated form of  \eqref{def:m0:rand} is a natural consequence of Theorem \ref{thm:IMC}, but if the variance of \linebreak $\widehat{m}_0 (r_1(p_1,U_1), \ldots, r_m( p_m,U_m))$ w.r.t. $U_1, \ldots, U_m$  is small it can be shown  that {$\mrand (p_1, \ldots, p_m) \approx \E_{U} \widehat{m}_0 (r(p,U))$, which is identical to $\mmean$ for  $\widehat{m}_0 \in \mathcal{F}_0$ (see \eqref{eq:identity:EU:mean} in  \ref{appendix:auxres}).

\begin{remark}	
In this section we  introduced three types of adjustments for null proportion estimators under discreteness.
They all address primarily the issue of super-uniformity and do so by \emph{individually}	shrinking the  $p$-values  to behave more like their uniform counterparts. Consequently, as these individual distributions become closer to the uniform distribution, a homogenizing effect on the whole set of distributions appears. Thus,  addressing super-uniformity in this way also counters heterogeneity (if it is present), but more as a byproduct.

While we may always expect	 both $\mmean$ and $\mrand$ to be beneficial both in the heterogeneous and homogeneous setting, the situation is slightly more complicated  for $\mresc$.  A necessary condition for $\mresc < \widehat{m}_0$, is that at least one of the $\nu^{\text{resc}}_i > \nu$, which is guaranteed to happen, whenever the $\nu^{\text{resc}}_i$-values are not identical, i.e. when the $F_i$'s are heterogeneous. In the homogeneous setting the behavior of $\mresc$ depends on the transformation function. As a first example take $g(u)=u$ which yields $\nu^{\text{resc}}_i>1/2$ so that the Pounds-Cheng estimator  improves by rescaling. This may not be the case for $g(u)=\ind{u>\lambda}$  if $\lambda$ is chosen from the support of  $F_1= \ldots = F_m$, i.e. for Storey's estimator we then have $\mresc = \widehat{m}_0$.
\end{remark}

\section{Results for simulated data}\label{ssec:simu:discrete}

In this section, we analyze how the discrete adjustments can improve the base estimators on simulated data. 
More specifically, we follow \cite{DDR2018} by simulating a two-sample problem in which a vector of $m= 500$ independent binary responses is observed for $N$ subjects in two groups. 
The goal is to test the $m$ null hypotheses $H_{0,i}$: '$p_{1i} = p_{2i}$', $i = 1,...,m$ where $p_{1i}$ and $p_{2i}$ are the success probabilities for the $i^{th}$ binary response in the two groups respectively. 
Thus, for each hypothesis $i$, the data can be summarized by a $2 \times 2$ contingency table, and we use (two-sided) Fisher's exact tests (FETs) for testing $H_{0i}$. 
{The $m=500$ hypotheses are split into three groups of size 
	$m_1$, $m_2$, and $m_3$ such that $m = m_1 + m_2 + m_3$.
	Then, the binary responses are generated as i.i.d Bernoulli of probability 0.01 ($\Bin(1,0.01)$) at $m_1$ positions for both groups,
	i.i.d $\Bin(1,0.10$) at $m_2$ positions for both groups, 
	and i.i.d $\Bin(1,0.10)$ at $m_3$ positions for one group
	and i.i.d $\Bin(1,0.40)$ at $m_3$ positions for the other group. 
Thus, null hypotheses are true for $m_1 + m_2$ positions, while they are false for $m_3$ positions. 
We set $m_1=m_2=150$ so that $m_0=300$ for the true number of null hypotheses (or accordingly $\pi_0=0.6$).}  
Note that the proposed methods can also be investigated for other types of discrete tests (e.g. Poisson) following \cite{durand2019discretefdr}.  
In the present work, we focus the simulation setting to FET as they appear also in all the real data results presented in Figures~\ref{fig:FETestim_IMPC} and Figure~\ref{fig:MultipleDiscretePlots}.



We perform $1\:000$ simulation runs in which we compute the base estimators  $\mStorey$ \eqref{def:m0:Storey} (with $\lambda = 0.5$) and $\mPCNew$ \eqref{eq:def:PC:new} 
with their three discrete variants given by $\mresc$ \eqref{eq:def:discrete:rescaled:estimator}, $\mmean$ \eqref{eq:def:discrete:adjusted:estimator} and $\mrand$ (see \eqref{def:m0:rand} and \eqref{def:m0:rand:MC} with $\NMC$ set to $1\:000$), presented in sections~\ref{sec:discretestimators}.
We present point estimation results for $\pi_{0}$ alongside power -- taken as the ratio between the number of true discoveries and the total number of signals -- and FDR estimates for the associated plug-in procedures.

Figure~\ref{fig:FETestim_base_vs_resc} displays the point estimation results for group size $N \in \{10, \dots, 100 \}$.
\cite{DMR2024} empirically show that this parameter can be seen as a proxy for the level of discreteness in the data: smaller values of $N$ tend to yield more discrete $p$-values as $N$ corresponds to the total marginal row count in the $m$ contingency tables. We can see that accounting for discreteness leads to considerable improvements for the base estimators $\mStorey$ and  $\mPCNew$  over the entire range of $N$ values. {More specifically, Figure ~\ref{fig:FETestim_base_vs_resc} shows that the efficiency gained by the discrete estimators is essentially the same for all three types of discreteness adjustments introduced in Section 3.  It also shows that  whereas the  base Pounds-Cheng estimator seems inferior to the base Storey estimator, its discrete variants yield a similar  performance. Adjusting for discreteness is especially useful  for small values of $N$ for which the base estimators are particularly inefficient, in many cases  providing estimation results above one. We mention  that the phenomenon that $\pi_0$-estimators may exceed one is not specific to the discrete setting  but may occur more generally in situations with sparse and/or weak signals (see e.g. Remark 16 in \cite{BR2009}).} 

\begin{center}
	\begin{figure}[H]  
		\center
		\includegraphics[width=.9\linewidth]{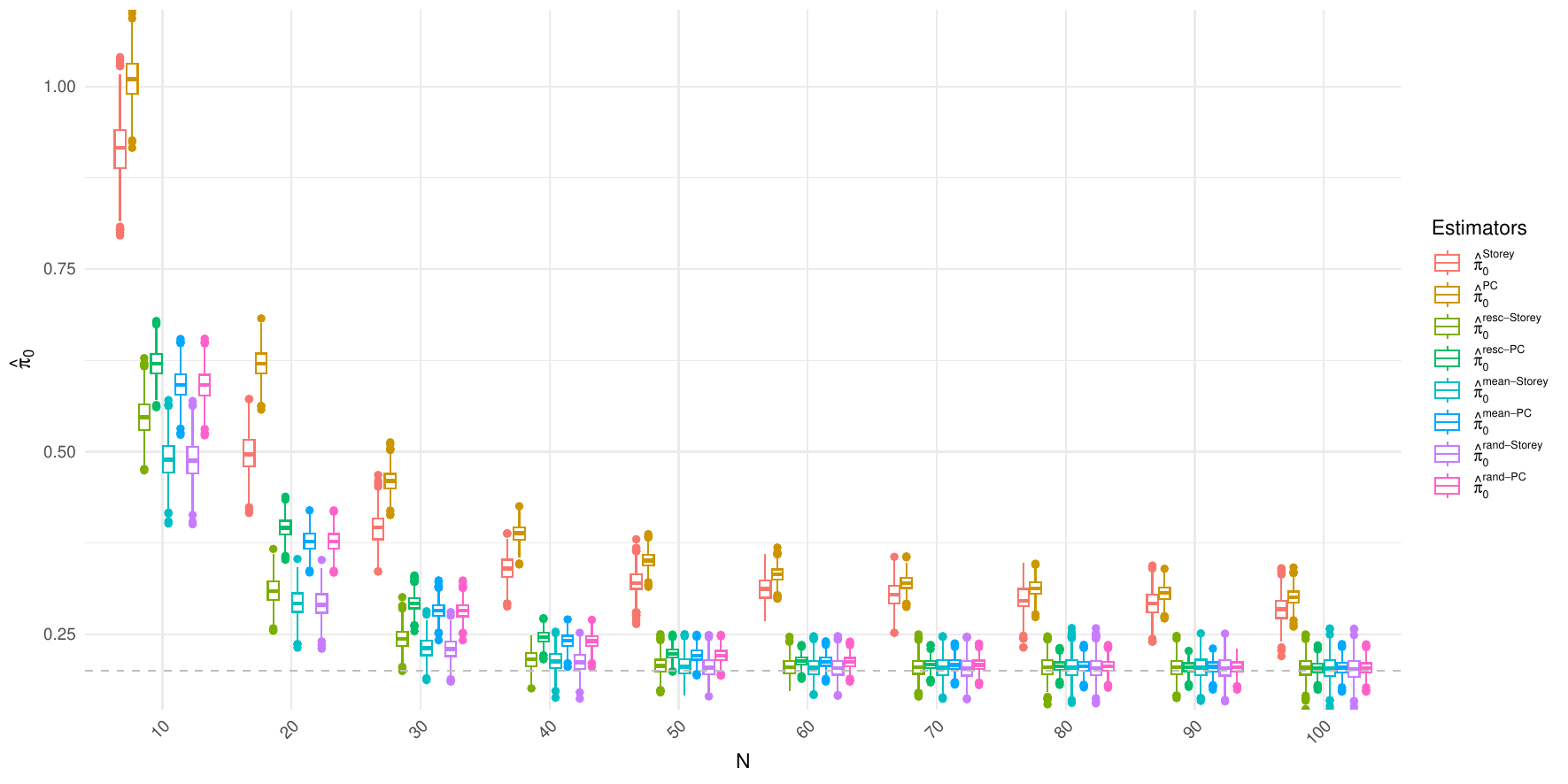}
		\includegraphics[width=.9\linewidth]{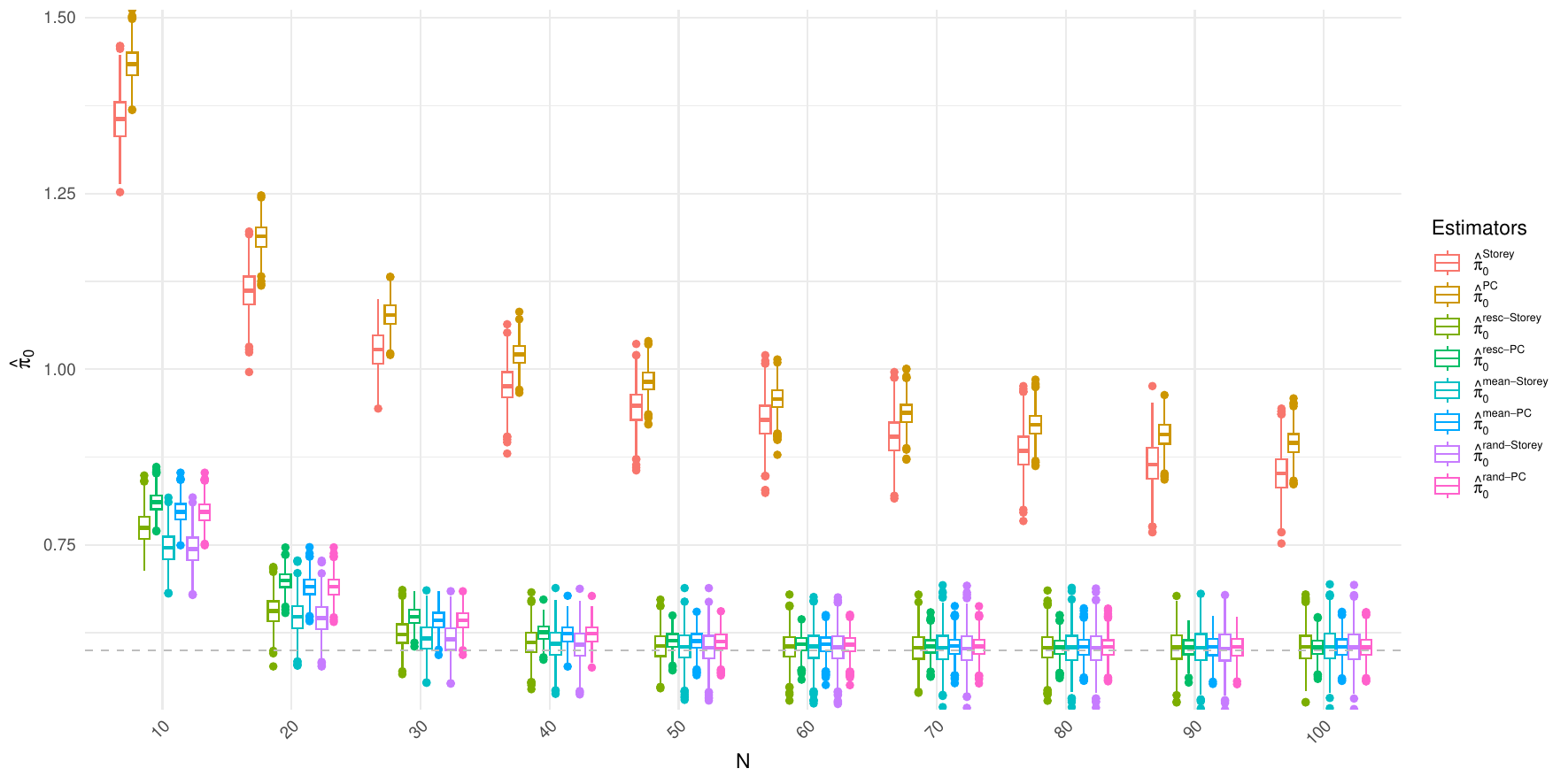}
		\includegraphics[width=.9\linewidth]{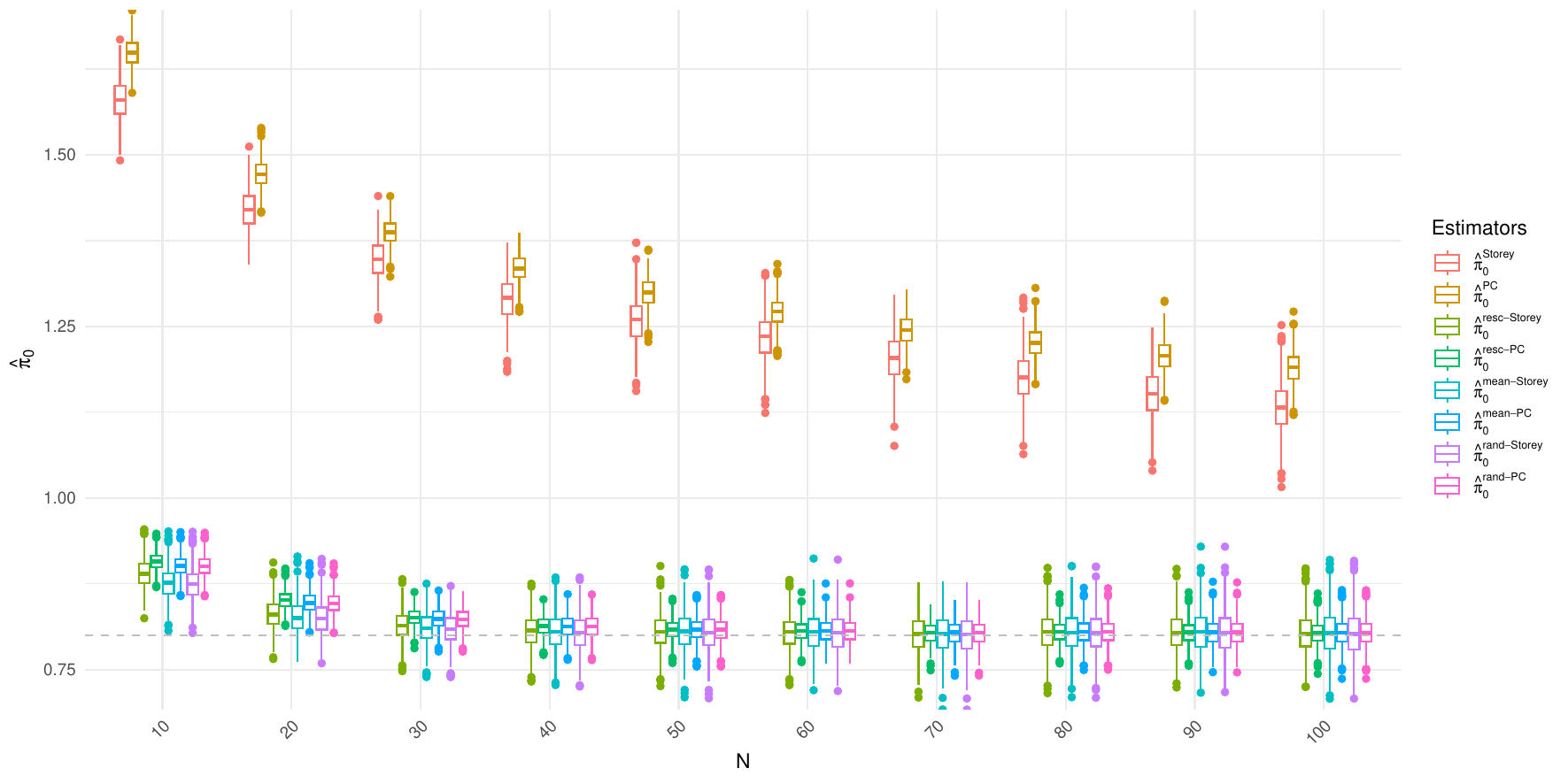}
		\caption{Boxplots of base estimators $\mStorey$, $\mPCNew$ and their discrete adjustments -- namely $\mresc$, $\mmean$ and $\mrand$ -- on simulated independent data. 
			The true value of null proportion in the three panels is set to $\pi_{0}=0.2,0.6$ and $\pi_{0}=0.8$ (from top to bottom) and indicated by the dashed gray line.}
		\label{fig:FETestim_base_vs_resc}
	\end{figure}
\end{center}

The improvements in estimation performance also carry over to improved power properties, as illustrated in the lower panel of Figure ~\ref{fig:disc-indep-fdr-power}, which presents results for each plug-in BH procedure at level $\alpha = 0.05$. 
\begin{center}
	\begin{figure}[hb]
			\center
		\includegraphics[width=.9\linewidth]{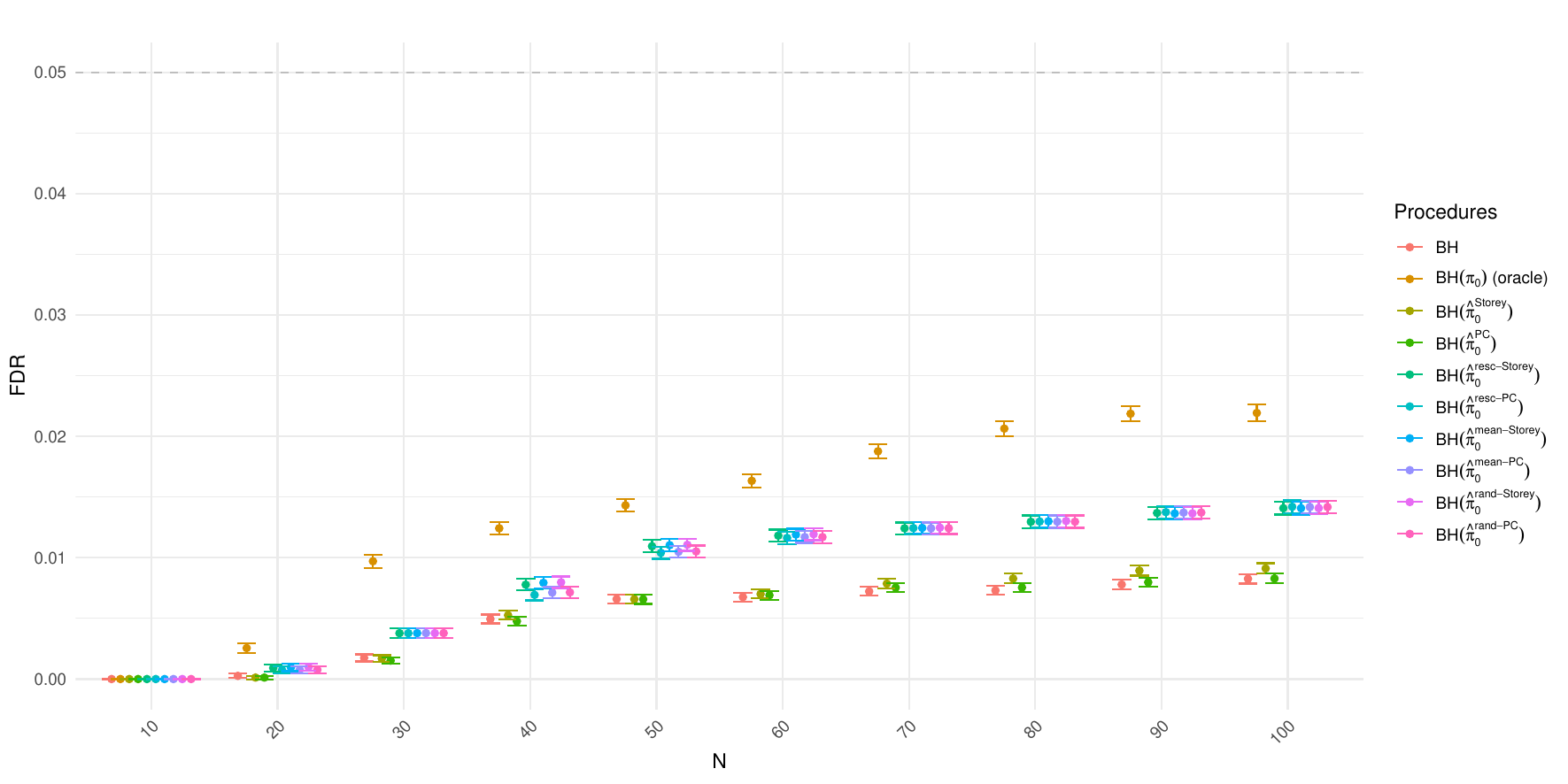}
		\includegraphics[width=.9\linewidth]{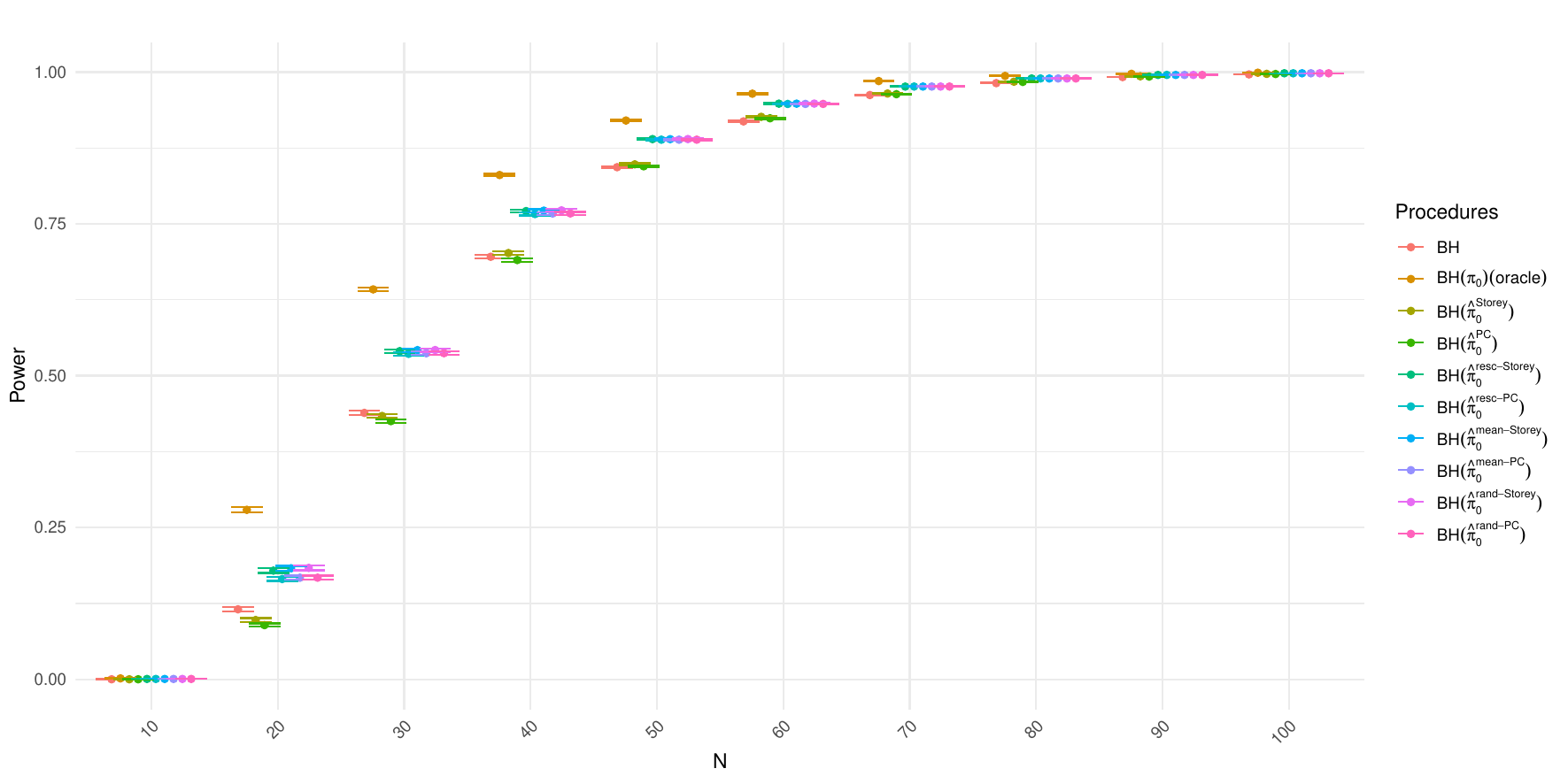}
		\caption{Confidence intervals (95$\%$) for FDR (upper panel) and power (lower panel) for plug-in BH procedures associated to $\mStorey$, $\mPCNew$ estimators, their discretized versions $\mresc$, $\mmean$ and $\mrand$, and the baseline BH and oracle BH  procedure on simulated independent data.} \label{fig:disc-indep-fdr-power}
	\end{figure}
\end{center}
For comparison, the figure also includes the performance of the unadjusted (raw) BH procedure and the oracle plug-in BH procedure (which uses the true value of $\pi_0$), representing  respectively what should be the least and most favorable procedures from a theoretical viewpoint. 
The upper panel of this figure shows that all procedures maintain FDR levels well below the nominal $0.05$ threshold. The improvement brought by discreteness-aware adaptivity is visible in the lower panel of Figure \ref{fig:disc-indep-fdr-power}. The upper panel of this figure highlights the remaining limitation due to the rejection step itself, as the oracle BH has also an FDR level of only 0.02.

\begin{center}
	\begin{figure}[h!]
		\centering
		\includegraphics[width=.9\linewidth]{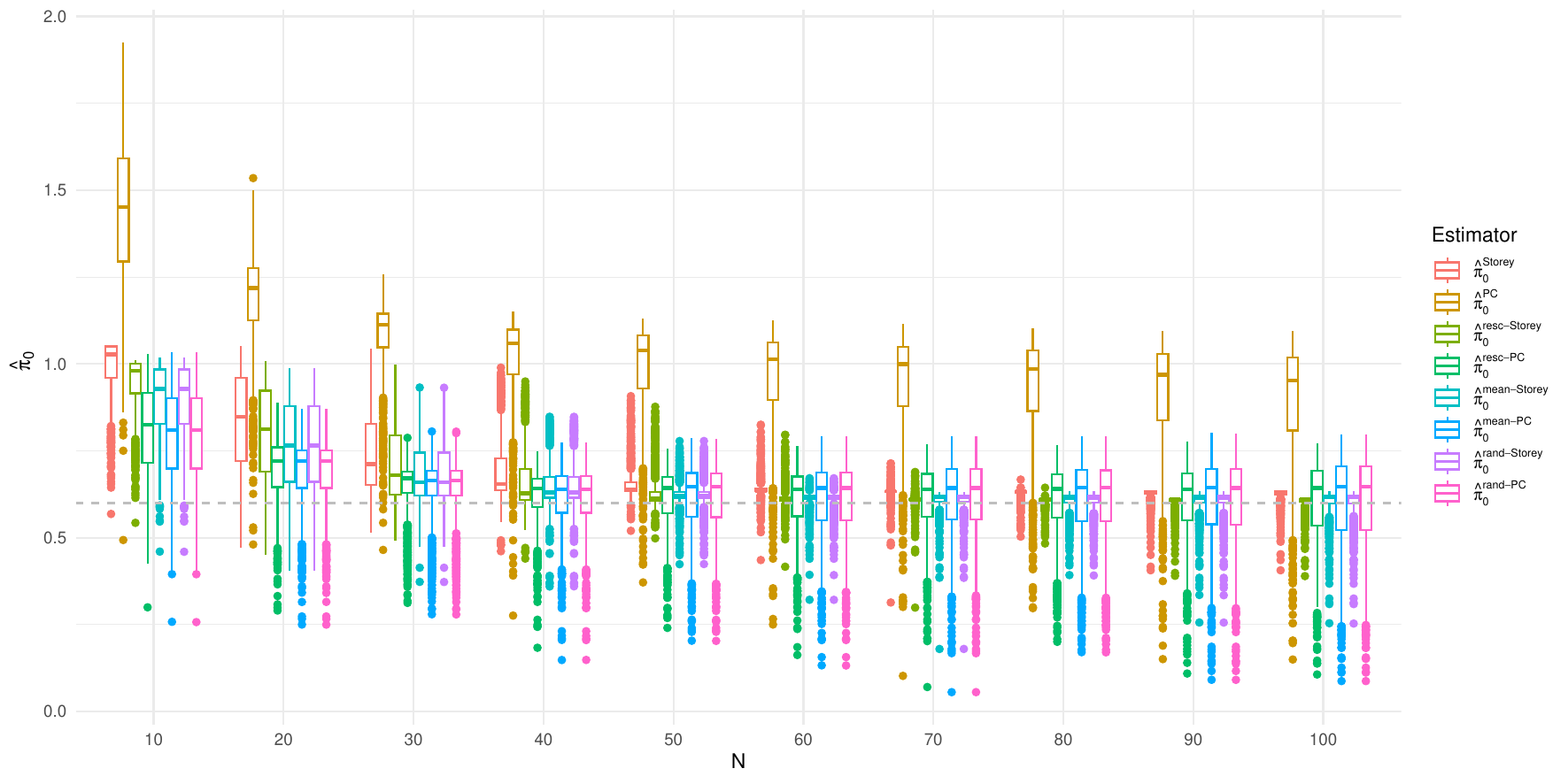}
		\caption{Boxplots of base estimators  $\mStorey$, $\mPCNew$ and their discrete adjustments -- $\mresc$, $\mmean$ and $\mrand$ -- on simulated dependent data.
			The true value of null proportion is set to $\pi_{0}=0.6$ and indicated by the dashed grey line.} \label{fig:FETestim_all_disc_dep}
	\end{figure}
\end{center}

Following, we conduct a similar analysis where we introduce positive dependence between the $p$-values. 
More specifically, we first simulate two multivariate uniform random vectors, each distributed according to the standard Gaussian copula with  an equicorrelation matrix with correlation coefficient set to $0.6$. In a second step, counts are generated by plugging these multivariate uniforms into the quantile function of a Bernoulli r.v. and the subsequent analysis proceeds as in the independent case described above.  
The true value of null proportion is again set to $\pi_{0}=0.6$, and we set $\lambda = \alpha (= 0.05)$ for $\mStorey$ following the recommendation of \cite{BR2009} for dealing with dependent $p$-values.

Figure~\ref{fig:FETestim_all_disc_dep} presents estimation results under this dependence structure, analogous to those shown in Figure~\ref{fig:FETestim_base_vs_resc} for the independent case. We observe that the discrete estimators remain robust to dependence: their median estimates closely match those obtained under independence. However, the increased variance -- expected in the presence of positive  correlation -- leads to greater variability, including more frequent underestimation (points falling below the dashed grey line).

Despite this, the performance of the plug-in BH procedures remains without concerns, as shown in the left panel of Figure~\ref{fig:disc-dep-fdr-power} as
all procedures maintain FDR control well below the nominal level $\alpha = 0.05$.
Similar to the independent case, this again, emphasizes the distinction of two levels of difficulty: even when the null proportion is underestimated, the procedure’s rejection capacity remains limited by the super-uniformity of discrete $p$-values. The dependence effect is again visible in the increased variability, as reflected by wider confidence intervals for both FDR and power. 

	\begin{figure}[H]
		\centering
		\includegraphics[width=0.9\linewidth]{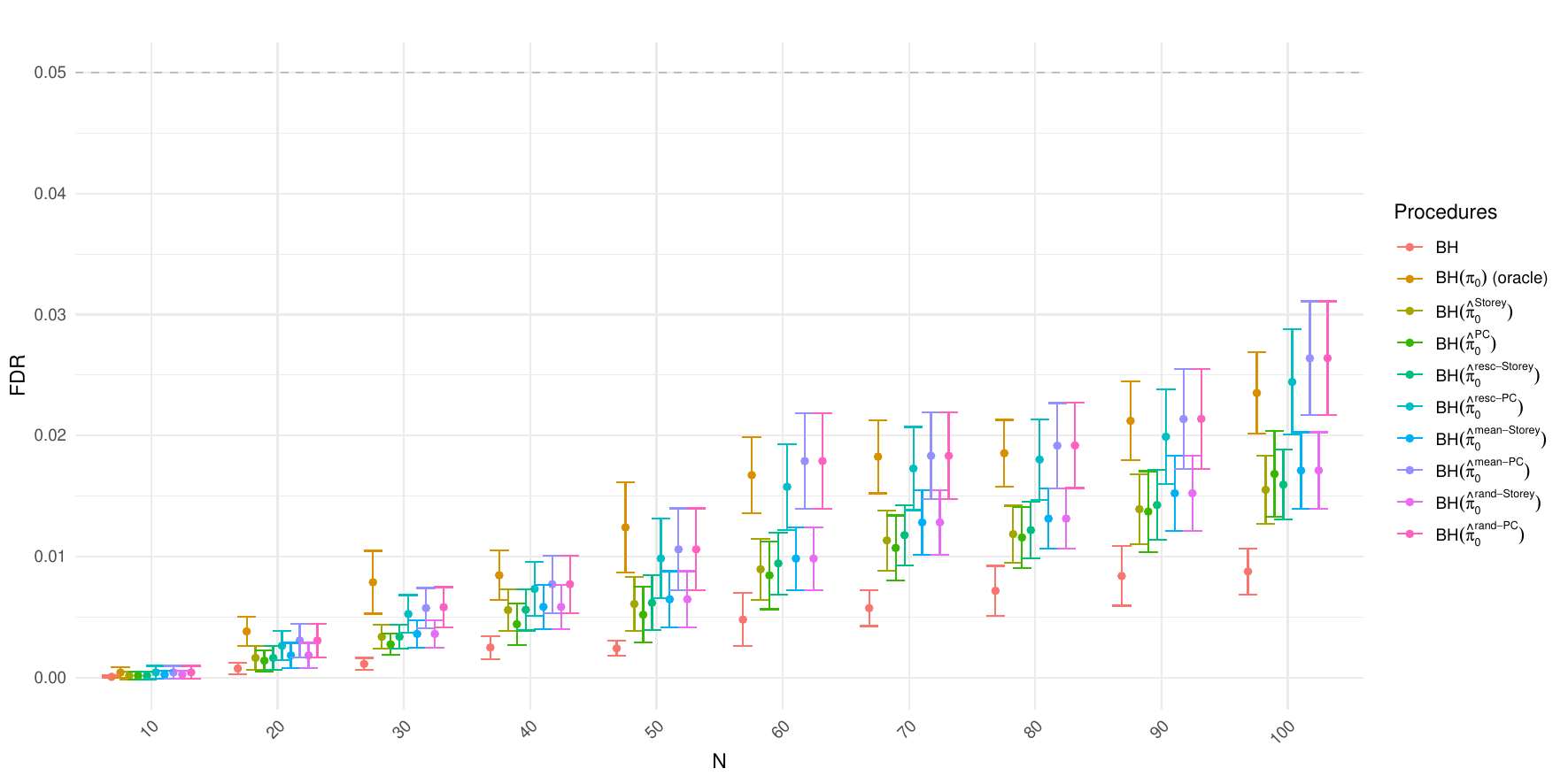}
		\includegraphics[width=0.9\linewidth]{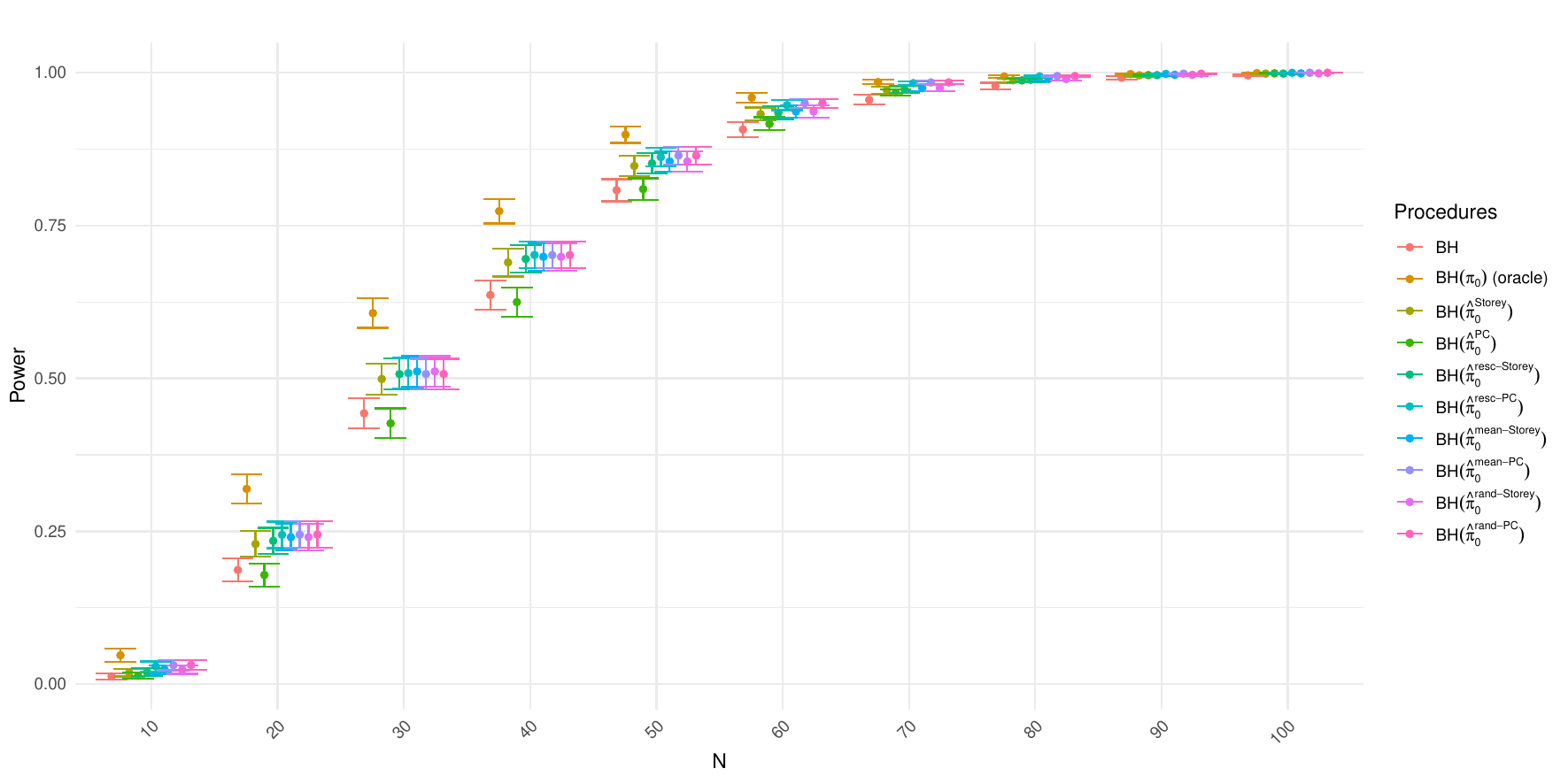}
		\caption{Confidence intervals (95$\%$) for FDR (top panel) and power results (lower panel) for plug-in BH procedures associated to $\mStorey$, $\mPCNew$ estimators, their discretized versions $\mresc$, $\mmean$ and $\mrand$, and the baseline BH and oracle BH (using the true $\pi_0$) procedure on simulated dependent data.} \label{fig:disc-dep-fdr-power}
	\end{figure}

\section{Real data analysis} \label{sec:real:data:analysis}
Finally, we illustrate the performance of base and discrete estimators on a real data-set (for more analyses on similar discrete data-sets we refer to  \ref{appendix:sec:additional:analyses:real:data}). We analyze data provided by the International Mouse Phenotyping Consortium (IMPC)  \citep{karp2017prevalence}, which coordinates studies on the genotype influence on mouse phenotype.  The IMPC data includes, for each of the $266952$ studied genes, the counts of normal and abnormal phenotypes thus providing multiple two by two contingency tables, which can be analyzed using (two-sided) FETs (for more details we refer to the R-package \texttt{DiscreteDatasets} and \cite{DMR2024}). In our analysis we focus on the first $m=3000$ genes in male mice  and use the same estimators used in the simulation study in Section \ref{ssec:simu:discrete}, i.e. with $\lambda=0.5$ for $\mStorey$ and its variants and $\NMC=1000$ for $\mrandMC$. 

Figure \ref{fig:FETestim_IMPC} displays the results of this analysis. Clearly, taking discreteness into account in $\pi_0$ estimation is  beneficial in terms of estimation and rejection performance. This is true for both base estimators and  the gains in efficiency are essentially the same for all three types of discreteness adjustments introduced in Section \ref{sec:discretestimators}. These findings are consistent with the analyses for simulated data in Section \ref{ssec:simu:discrete} and similar results also hold true for other  data-sets analyzed in  \ref{appendix:sec:additional:analyses:real:data}. 
\begin{figure}[htbp]
    \centering
    \includegraphics[width=.9\linewidth]{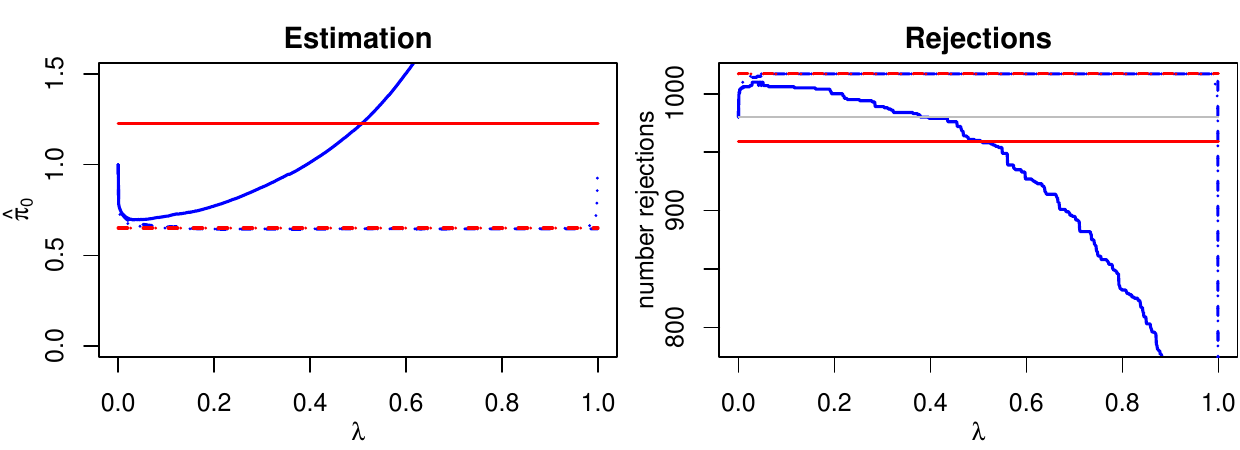}
    \caption{$\pi_{0}$-estimators (left panel) and number of discoveries for the plug-in BH procedure (right panel) on the IMPC data-set. Blue lines represent versions of the Storey estimator, red lines represent versions of the PC estimator. For both estimators the solid lines represent the base (continuous) estimator, the dashed lines represent the rescaled \eqref{eq:def:discrete:rescaled:estimator} and the dotted lines the conditional mean \eqref{eq:def:discrete:adjusted:estimator} modifications for discrete data. The thin grey  line in the right panel represents the number of discoveries by the (non-adaptive) BH procedure. The results for the randomized estimator \eqref{def:m0:rand} are practically identical with those for the conditional mean estimator  and have therefore been omitted from the graphs for better readability.}
    \label{fig:FETestim_IMPC}
\end{figure}

Depending on the type of discreteness and the amount of signal contained in the data, the  gain in efficiency may be considerable. 
For Storey's estimator, which requires choosing an adequate value of the tuning parameter $\lambda$, the discrete adaptations also exhibit more stable behavior over the range of tuning parameter values.  
On another note, whereas the base PC estimator is clearly outperformed by the base Storey estimator for many values of $\lambda$, the discrete versions of these two estimators exhibit very similar behavior.

\section{Some remarks on the choice of transformation function}\label{sec:outlook}

The estimators introduced  in this paper are essentially sums of  $p$-values transformed by some function $g \in {\mathcal{G}}$. The class ${\mathcal{G}}$ allowing great freedom, a natural question to ask is how to choose $g$ well. A comprehensive answer to this  question is beyond the scope of this paper, but in this section we   sketch some paths towards answers  that may be useful for further investigations. First we describe two approaches in the discrete data setting for obtaining transformations with minimal bias. Then we develop a method for minimal MSE in the continuous case. In all cases, numerical methods are needed to solve the resulting optimization problems.

\subsection{Towards estimators with minimal bias in the discrete setting}\label{ssec:mse:optimal:discrete}
As a starting point we consider a single discrete $p$-value as in the beginning of Section \ref{sec:discretestimators}, i.e. $p \sim F_0$ under $H_0$ with support   $\mathcal{S}=\{0<s_1 < s_2 < \ldots < s_L=1\}$. Clearly, for any non-decreasing function $g:[0,1] \rightarrow [0,1]$ the distribution of the random variable $g(p)$ is completely determined by its restriction to $\mathcal{S}$ so that a.s. 
\begin{align}
	g(p) &= \sum_{j=1}^{L} w_j \cdot \ind{p \ge s_{j}} =: g_{\bw}(p) \label{eq:def:g:w}\\
\intertext{where $w_1 = g(s_1)$ and $	w_i = g(s_i)-g(s_{i-1})$ . Conversely, any vector}
\bw &\in \mathcal{W} = \{\bw=(w_1, \ldots, w_L) \in [0,1]^L | \|\bw\|_1 \le 1\}\notag
\end{align}
defines a proper function $g_{\bw} \in {\mathcal G}$ as in \eqref{eq:def:g:w}, so that in our search for good candidate transformations $g \in {\mathcal G}$ we can w.l.o.g. restrict ourselves to the sub-class $\{g_{\bw} | \bw \in \mathcal{W} \}$. For the expectation of $g_{\bw}$ we have (with slight notational abuse)
\begin{align}
	\nu = \nu_{\bw} &= \E_0 g_{\bw}(p) = \sum_{j=1}^{L} w_j \cdot P_0(p \ge s_{j}) =: \bw^T \cdot \widetilde{F_0}(\mathcal{S}),  \label{eq:def:nu:w}
\end{align}
where $\widetilde{F_0}(\mathcal{S})=(1,1-s_1,1-s_2, \ldots, 1-s_{L-1})$.
Since we are dealing with multiple $p$-values $p_1, \ldots, p_m$ each $p$-value may have its own null distribution function $F_{01}, \ldots, F_{0m}$ with support $\mathcal{S}_1,\ldots,\mathcal{S}_m$. Defining $g_1=g_{\bw_1}, \ldots, g_m=g_{\bw_m}$ for weight vectors $\bw_1 \in \mathcal{W}_1, \ldots, \bw_m \in \mathcal{W}_m$ as above, any estimator $\widehat{m}_0 \in \mathcal{F}$ (see \eqref{eq:def:class:estimators}) can therefore be expressed as
\begin{align}
	\widehat{m}_0 = \widehat{m}_0 (\bw_1, \ldots, \bw_m) &= \frac{1}{\min(\nu_{\bw_1}, \ldots, \nu_{\bw_m})} + \sum_{i=1}^{m} \frac{g_{\bw_i}(p_i)}{\nu_{\bw_i}}, \label{eq:def:mZero:discrete:general}
\end{align}
and by a similar reasoning as in the proof of Proposition \ref{prop:plugin:discrete:control:general:g}	$\widehat{m}_0$ provides plug-in FDR control.}

Suppose now that we are interested in finding estimators defined by  \eqref{eq:def:mZero:discrete:general} with minimal bias. Similar to \eqref{eq:bias:m0:general} we obtain
\begin{align}
\bias(	\widehat{m}_0) &= \E \widehat{m}_0 -m_0 = \frac{1}{\min(\nu_{\bw_1}, \ldots, \nu_{\bw_m})} + \sum_{i \in \nullset } \left(\frac{\E_0 g_{\bw_i}(p_i)}{\nu_{\bw_i}}-1\right) + \sum_{i \in \altset } \frac{\E_1 g_{\bw_i}(p_i)}{\nu_{\bw_i}} \notag\\
&= \frac{1}{\min(\nu_{\bw_1}, \ldots, \nu_{\bw_m})}  + \sum_{i \in \altset } \frac{\E_1 g_{\bw_i}(p_i)}{\nu_{\bw_i}}, \label{eq:bias:alternatives:hetero}
\end{align}
where the 'null-bias' disappears in  \eqref{eq:bias:alternatives:hetero} due to the definition of $\nu_{\bw}$ in \eqref{eq:def:nu:w}, and $\E_0, \E_1$ denote  expectation respectively under the null and the alternative. Equation \eqref{eq:bias:alternatives:hetero} implies that the bias depends on the distribution of $p$-values under the alternatives so that merely knowing $F_{01}, \ldots, F_{0m}$ will generally not be sufficient for constructing  estimators with low bias. Thus, we need to make some additional assumptions on the data generating process. If we do this in an a-priori sense (i.e. without seeing the data) plug-in FDR control is still guaranteed. To fix ideas, let us assume a two-groups mixture model, i.e. $p_1, \ldots, p_m$ are independent with 
\begin{align}
	p_i &\sim \pi_0 F_{0i} + (1-\pi_0)F_{1i} \label{eq:def:hetero:mixtures:model}
\end{align} with known $\pi_0$ and  $p$-value distributions under the alternatives $F_{11}, \ldots, F_{1m}$. In this model we have 
\begin{align}
	\bias(	\widehat{m}_0)  &=  \frac{1}{\min(\nu_{\bw_1}, \ldots, \nu_{\bw_m})}  + (1-\pi_0)\sum_{i =1}^m \frac{\E_1 g_{w_i}(p_i)}{\nu_{\bw_i}} \notag \\
	&=  \frac{1}{\min(\bw_1^T \cdot \widetilde{F_{01}}(s_1), \ldots, \bw_m^T \cdot \widetilde{F_{0m}}(s_m))}  + (1-\pi_0)\sum_{i =1}^m \frac{\bw_i^T \cdot \widetilde{F_{1i}}(s_i)}{\bw_i^T \cdot \widetilde{F_{0i}}(s_i)},
\end{align}
so that minimizing the bias is equivalent to solving the following optimisation problem (for brevity we set $x_i=\widetilde{F_{0i}}(s_i), y_i=\widetilde{F_{1i}}(s_i)$):
\begin{mini}
	{\bw_1, \ldots,\bw_m}{\frac{1}{\min(\bw_1^T \cdot x_1, \ldots, \bw_m^T \cdot x_m)} + (1-\pi_0)\sum_{i =1}^m \frac{\bw_i^T \cdot y_i}{\bw_i^T \cdot x_i},}{}{}
	\addConstraint{\bw_1, \ldots, \bw_m}{\geq 0}
	\addConstraint{\|\bw_1\|_1 ,\ldots, \|\bw_m\|_1 }{\le 1}. \label{eq:optimisation:bias:discrete:general}
\end{mini}
At first sight, this appears to be a complicated, non-standard optimization program. However, it can be shown that this program can be transformed into a quadratically constrained linear program (see Proposition \ref{prop:Tobias}). The transformed program is still NP-hard, but with state-of-the-art solvers (e.g.  \cite{gurobi2026}) a near-optimal solution can be found in appropriate time, yielding  near-optimal transformations $g_{\bw_1}, \ldots, g_{\bw_m}$.
%
	
The previous approach yields the best estimator in the whole class $\mathcal{F}$. If we are only interested in the smaller class of Storey-type estimators, there is even a much simpler solution. By taking $g(p) =1\{p > \lambda\}$, i.e. the \emph{same} transformation for all $p$-values, we obtain in the two-groups mixture model \eqref{eq:def:hetero:mixtures:model}   
\begin{align}
	\bias(	\widehat{m}_0)=\bias(	\widehat{m}_0 (\lambda))  &=  \frac{1}{\min(1-F_{01}(\lambda), \ldots, 1-F_{0m}(\lambda))}  + (1-\pi_0)\sum_{i =1}^m \frac{1-F_{1i}(\lambda)}{1-F_{0i}(\lambda)}, \label{eq:Store:discrete:bias:mixture} 
\end{align}
where -- due to the discreteness of the involved distributions -- the r.h.s. of  \eqref{eq:Store:discrete:bias:mixture} is a step function with jumps occurring only at points contained in the joint support $\mathcal{S}=\mathcal{S}_1 \cup \ldots \cup \mathcal{S}_m$. Therefore,  it is sufficient to evaluate  $\bias(	\widehat{m}_0 (\lambda)) $ only on the finite set $\mathcal{S}$ without the need of more sophisticated optimisation algorithms as in the  general approach outlined above.

A detailed investigation of the two approaches exceeds the scope of this paper and is therefore left for future work. We also leave the extension to MSE-optimal estimators for future work but nevertheless present  a related result for the continuous case in the next section.

\subsection{Towards $\mse$-optimal estimators in the continuous setting}\label{ssec:mse:optimal:homogeneous}
When the tests statistics are continuous, i.e. $p_i \sim U(0,1)$ under the nulls, we can extend the approach described above to obtain approximately $\mse$-optimal estimators. Our starting point is the class  of estimators with
\begin{align}
	\widehat{\pi}_0&= \widehat{\pi}_0 (g) = \frac{1}{m \nu(g)}\left( 1 + \sum_{i=1}^m  g(p_i)\right) \label{eq:def:generalized:Storey}
\end{align}
where $g \in {\mathcal{G}}$. For these estimators the bias, variance and MSE can be obtained as follows.

\begin{proposition}\label{prop:bias:variance:MSE}
For $\widehat{\pi}_0$ as in \eqref{eq:def:generalized:Storey} we have

\[
\begin{aligned}
\bias(\widehat{\pi}_0)
&= \frac{1}{m\nu}
 +(1-\pi_0)\frac{\E_1 g(p)}{\nu},
\qquad
\var(\widehat{\pi}_0)
= \frac{1}{m\nu^2}
\left[
\pi_0\var_0 g(p)
+(1-\pi_0)\var_1 g(p)
\right].
\end{aligned}
\]
and,
\begin{align}
\mse(\widehat{\pi}_0)
&=
\frac{1}{m\nu^2}
\Biggl[
\frac{1}{m}
-\pi_0\nu^2
+2(1-\pi_0)\E_1 g(p)
\\
&\qquad
+(m-1-m\pi_0)(1-\pi_0)(\E_1 g(p))^2
+\pi_0\E_0 g(p)^2
+(1-\pi_0)\E_1 g(p)^2
\Biggr].
\label{eq:MSE:general}
\end{align}
\end{proposition}

%

%

In what follows, our goal is to find a $\mse$-optimal transformation $\widetilde{g} \in {\mathcal G}$ for estimators as in \eqref{eq:def:generalized:Storey} so that
\begin{align}
	\mse(\widehat{\pi}_0(\widetilde{g})) &= \min_{g \in {\mathcal G}} \mse(\widehat{\pi}_0(g)) \label{eq:optimisation:problem:general}
\end{align}
in the oracle case, i.e. where the data generating mechanism for $p_1, \ldots, p_m$ is known. Ideally, we would like to find an analytical expression for $\widetilde{g} \in {\mathcal G}$, depending on the model parameters but we do not pursue  this aim further here. Instead we show that it is relatively easy to find an approximate (numerical) solution to the optimization problem \eqref{eq:optimisation:problem:general} by approximating the complex function class ${\mathcal G}$ by a smaller and simpler class of convex combinations of  Storey-type indicator functions. More specifically, for  $K \in \N$ define sets of grids and possible weights
\begin{align*}
	\Lambda &= \{ \blambda=(\lambda_1, \ldots, \lambda_K) \in [0,1)^K | 0=\lambda_1 < \lambda_2 < \ldots < \ldots \lambda_K <1\},\\
	{\mathcal W} &= \{\bw=(w_1, \ldots, w_K) \in [0,1]^K | \|w\|_1 \le 1\}.
	\intertext{For a fixed $\blambda \in \Lambda$ we introduce the approximating class of transformation functions}
	{\mathcal G} (\blambda)&= \{g_{\bw,\blambda}:t \mapsto \sum_{k=1}^K w_k \cdot 1_{(\lambda_k,1]}(t) | \bw \in {\mathcal W} \}
	\intertext{which can be seen as the convex hull over certain Storey-type indicator transformations. The heuristic  behind this approach is that any reasonable solution $\widetilde{g} \in {\mathcal G}$ of \eqref{eq:optimisation:problem:general} should be well approximable by some  $g^{\ast} \in {\mathcal G} (\blambda)$ if the grid $\blambda$ is fine enough. Clearly, ${\mathcal G} (\blambda) \subset {\mathcal G}$ and this defines a new sub-class of estimators as in \eqref{eq:def:generalized:Storey}, i.e. for $\bw \in {\mathcal W} $}
	\piZeroConv(\bw)&=\frac{1}{m \nu_{\bw,\blambda}}\left( 1 + \sum_{i=1}^m  g_{\bw,\blambda}(p_i)\right),
\end{align*}
where $\nu_{\bw,\blambda}= \E g_{\bw,\blambda}(U)$.

The next result shows that in a two-groups model the $\mse(\piZeroConv(w))$ can be expressed in terms of a simple quadratic form. This will allow us to use standard quadratic programming methods to find (nearly) optimal solutions below.
\begin{proposition} \label{prop:MSE:quadratic:oracle}
	Assume a two-groups model, i.e.  $p_1, \ldots, p_m $ independent with
	\begin{align}
		p_i \sim \pi_0 U(0,1) + (1-\pi_0)F_1 \label{eq:def:two:groups:model}
	\end{align}	
	and let $K \in \N$ and $\blambda \in \Lambda$ be fixed.	Then there exists a vector $d=d(\blambda) \in \R^K$ and a symmetric matrix $C=C(\blambda) \in \R^{K \times K}$ so that for any $\bw \in {\mathcal W} $ it holds
		\begin{align}
			m\nu^2 \cdot \mse(\piZeroConv(\bw)) &=  \left[ \frac{1}{m} - \pi_0 \nu^2 +  \bw^T \cdot d + \frac{1}{2} \bw^T \cdot C \cdot \bw  \right], \label{eq:mse:two:groups:matrix}
		\end{align}
		where $\nu=\nu_{\bw,\blambda}= \bw^T \cdot (1- \blambda)$.
\end{proposition}

In order to find an approximately optimal transformation function $g_{{\bw}^{\ast},\blambda} \in {\mathcal G} (\blambda)$ we need to find an optimal weight ${\bw}^{\ast} \in {\mathcal W}$ so that 
\begin{align}
	\mse (\piZeroConv({\bw}^{\ast})) &= \min_{w\in {\mathcal W} } \mse (\piZeroConv(\bw)).\label{eq:optimisation:problem:convex}
\end{align}
For this, we procede in two steps:

\begin{enumerate}
	\item For arbitrary but fixed $\nu \in (0,1]$ find ${\bw}^{\ast}= {\bw}^{\ast}(\nu)$, so that $		\mse(\piZeroConv ({\bw}^{\ast}) )$ is minimal.	Proposition \ref{prop:MSE:quadratic:oracle}  implies that minimizing $\mse (\piZeroConv(\bw))$ is equivalent to minimizing $w^T \cdot d + \frac{1}{2} w^T \cdot C \cdot w$ under the constraint  $\bw^T \cdot (1- \blambda)= \nu$ which leads to the following collection of  quadratic programming problems \eqref{eq:optimisation:problem:convex:fixed:nu}$_{\nu \in (0,1]}$: \\
	For each $\nu \in (0,1]$ find  ${\bw}^{\ast} = {\bw}^{\ast}(\nu) \in [0,1]^K$ with
	\begin{mini*}
		{\bw}{\frac{1}{2} \bw^T C \bw + d^T \bw ,}{}{}\label{eq:optimisation:problem:convex:fixed:nu} 
		\addConstraint{\bw}{\geq 0}
		\addConstraint{\bw^T \cdot (1- \blambda)}{= \nu} \tag{P$_\nu$}
		\addConstraint{\|\bw\|_1}{\le 1.}
	\end{mini*}
	\item The collection \eqref{eq:optimisation:problem:convex:fixed:nu}$_{\nu \in (0,1]}$ of optimization problems yields a collection $({\bw}^{\ast}(\nu))_ {\nu \in (0,1]}$ of optimal weights. For each ${\bw}^{\ast}(\nu)$  of \eqref{eq:optimisation:problem:convex:fixed:nu}  we have by equation \eqref{eq:mse:two:groups:matrix}
	\begin{align*}
		\mse(\piZeroConv ({\bw}^{\ast}(\nu))) &= \frac{1}{m\nu^2}\left[ \frac{1}{m} - \pi_0 \nu^2 +  {\bw}^{\ast}(\nu)^T \cdot d + \frac{1}{2} {\bw}^{\ast}(\nu)^T \cdot C \cdot {\bw}^{\ast}(\nu)  \right],
		\intertext{and therefore we can find the global minimum in \eqref{eq:optimisation:problem:convex} by}
		\mse(\piZeroConv ({\bw}^{\ast}) ) &= \min_{\nu \in (0,1]}\mse(\piZeroConv ({\bw}^{\ast}(\nu))).
	\end{align*}
\end{enumerate}
Equation \eqref{eq:mse:two:groups:matrix} suggests that the quantity $\nu$ can be viewed as a tuning parameter for $\piZeroConv$.	For the classical Storey-estimator $\widehat{\pi}^{Storey}_0$ with tuning parameter $ \lambda^{Storey} \in [0,1)$ we have  $\nu_{\bw,\blambda}=1 - \lambda^{Storey}$ for $\bw=(0,1)$ and $\blambda=(0,\lambda^{Storey})$. More generally,  large values of $\nu$ lead to estimators with large bias and small variance and vice versa for small values of $\nu$. Thus, the parameter $\nu$ can be interpreted as a value that determines/drives  the bias/variance trade-off. In practical applications we let $\nu$ run over a suitably fine grid of $\nu$-values in step 2. of the above algorithm. While this approach is heuristic, we may still expect that if the grid $\lambda$ is fine  enough the resulting new transformation function $\gNewStar = g_{{\bw}^{\ast}, \lambda}$ will approximate the optimal $\widetilde{g}$ of \eqref{eq:optimisation:problem:general} well. The following  example illustrates the behavior of $\gNewStar$.

 \begin{figure}[htbp]
	\centering
	\includegraphics[width=1\linewidth]{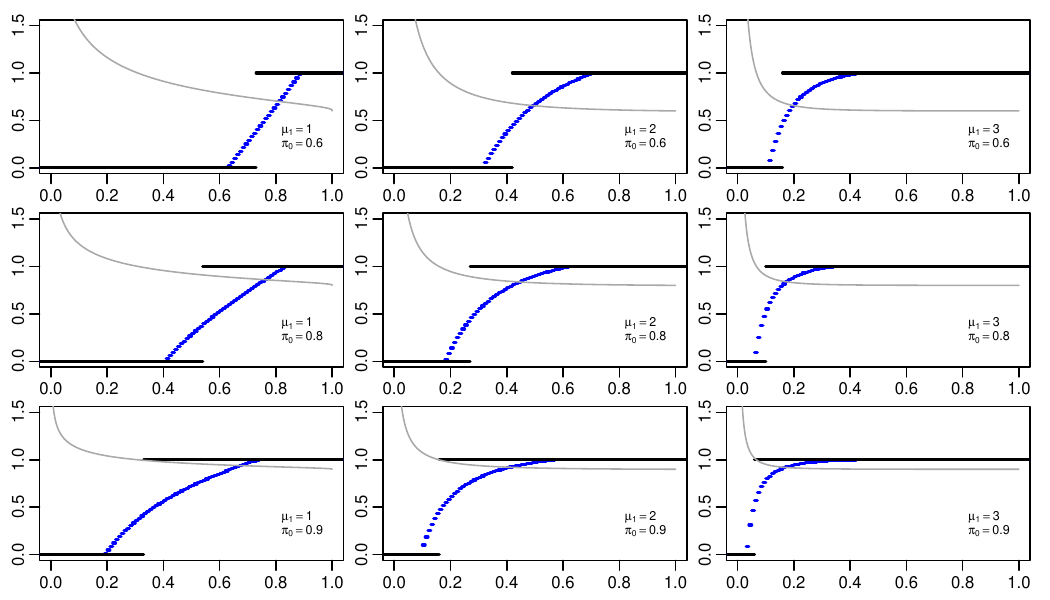}
	\caption{Graphs of  $\gNewStar$ for $\blambda = \left(0,1/100, \ldots,99/100\right)$ (blue curve), $\gStoreyStar$ (black curve) and the mixture density  (grey curve) with $(\mu_1,\pi_0) \in \{1,2,3\} \times \{0.6,0.8,0.9\} $ in Example \ref{example:gaussian:onesided:optimal:mse}.}
	\label{fig:OptimalNewStoreyMSE}
\end{figure}

\begin{example} \label{example:gaussian:onesided:optimal:mse}
We consider a one-sided Gaussian  mixture model, i.e.  $p_1, \ldots, p_m $ independent as in \eqref{eq:def:two:groups:model} where  $F_1(t) = \overline{\Phi}(\overline{\Phi}^{-1}(t)-\mu_1)$, i.e. $\mu_1$ is the effect size under the alternatives. We take $m=300$, $\blambda = \left(0,1/100, \ldots,99/100\right)$ with model parameters $\pi_0 \in \{0.6,0.8,0.9\}$ corresponding to small, medium and high probabilities of null hypotheses and $\mu_1 \in \{1,2,3\}$ corresponding to weak, moderate and strong signals. Figure \ref{fig:OptimalNewStoreyMSE} displays the functions $\gNewStar$ and the optimal Storey-type transformation $\gStoreyStar$, which is determined  by numerically minimizing \eqref{eq:MSE:Storey}, see Remark \ref{prop:bias:variance:MSE:Storey}, as well as the mixture density of model \eqref{eq:def:two:groups:model}.

Both $\gNewStar$ and  $\gStoreyStar$ tend to increase as the signal strength increases. This seems plausible as in the extreme case of $F_1 \sim \delta_{0}$ (Dirac-Uniform configuration) Proposition \ref{prop:bias:variance:MSE} suggests that $g$ should be chosen as $1_{(0,1]} $. As $\pi_0$ increases, $g$ should also increase since we want to estimate $\pi_0$. Whereas the Storey transformation measures the contribution of individual $p$-values to the $\pi_0$ estimate in a binary way, the new transformation 'interpolates' the Storey transformation, thereby smoothing the influence of individual $p$-values.
\end{example}

The above example shows in a Gaussian setting that (approximately) $\mse$-optimal transformations map $p$-values in a smoother way than the indicator functions underlying Storey's estimator. These transformation functions depend however on  model assumptions like  \eqref{eq:def:two:groups:model}, thus exposing the resulting estimators to the risk of poor performance due to model or parameter misspecification.  This suggests that more non-parametric  smooth transformations like kernel functions (see \cite{Neuvial2013}) could lead to useful estimators. We leave this for future investigations.

\section{Discussion} \label{sec:conclusion}


In this paper, we introduced  a simple and flexible class of $m_{0}$-estimators with  guaranteed plug-in FDR control. 
We placed particular emphasis on mitigating the conservativeness inherent in the statistical analysis of discrete data sets. 
To this end, we presented three approaches that adapt classical estimators to the discrete paradigm. 
We proved these methods to be uniformly more powerful than their non-discrete counterparts and  illustrated the gains in efficiency on both real and simulated data.

In Section~\ref{sec:discretestimators} we illustrated how information on the null distribution functions of discrete $p$-values can be used to obtain more efficient $m_0$-estimators. 
This information can be seen as a special case of auxiliary covariates, for which it is well-known that their incorporation into multiple testing procedures, e.g. by weighting, can be highly beneficial (see \cite{IHW,Guillermo}). 
We would like to mention that our methods for $m_0$-estimation  are not limited to the special case of discrete $p$-values, but should be able to accommodate other types of heterogeneity as well. 

Our analyses of real and simulated data suggest that for a given base estimator all adjustment methods yield similar performance. Therefore, we are reluctant to recommend a 'one-size-fits-all' adjustment, as no single approach can adequately accommodate the wide variety of data structures and scientific settings encountered in practice. Instead, we view our methods as a flexible toolbox that enables data analysts to explore different strategies for handling discreteness. From a practical viewpoint, the most appropriate choice may also depend on established practices within a given scientific community. For instance, the expected randomization method may be particularly easy to communicate to researchers already familiar with randomization-based approaches, whereas, for scientists accustomed to mid p-values, the conditional mean method may appear as a natural extension. When addressing a broader scientific audience, the rescaling approach—arguably requiring the least statistical background—may provide the most accessible option.

Beyond the discrete paradigm, numerous possibilities for extensions and further investigations exist.
For example, we derived a simple variant of the Pounds–Cheng estimator with FDR guarantees. 
More broadly, one could explore smooth $m_0$-estimators within our class, e.g. using kernel-based transformations \citep{Chen1999,Neuvial2013}, which might stabilize estimators such as Storey’s as pointed out by \citet{PC2006}.

It is worth noting that our work is related to that of \cite{heesen2016dynamic} who split the unit interval into an estimation region on which an estimator of $m_{0}$ is constructed and a rejection region on which the BH procedure is run. 
Thus, these estimators do not use all available $p$-values, in contrast to our approach. 
\cite{heesen2016dynamic} derive a general sufficient criterion, similar to Theorem~\ref{thm:IMC}, for finite sample plug-in FDR control which they apply to Storey-type estimators and histogram-type estimators  (see \cite{Macdonald2019}).  In contrast to our approach, the transformation  applied to $p$-values need not be monotone, however it is unclear whether e.g. smooth functions fit into this framework.  Their approach also accommodates ``dynamization'', which allows data-dependent tuning of parameters, see \cite{Macdonald2019, Gao2024}. 
An interesting question for future work could be whether such ideas can be integrated into our class of estimators.

Constructing multiple testing procedures for discrete data that provide finite sample plug-in FDR control is challenging.  
In this paper, we make some progress by obtaining improved discrete estimators for $m_0$.
While using these discrete estimators in the plug-in BH procedure provide more power than using classical estimators, it is still not ideal because the discreteness is ignored in the rejection stage of the procedure.
\cite{DDR2018} propose discrete variants of the standard (i.e. non plug-in) BH procedure.  
They also sketch a possible plug-in method based on combining this procedure with estimators of $m_0$,  but caution that it comes without mathematical guarantees. 
Thus, as \cite{Macdonald2019} pointed out, it still remains an open problem to develop procedures that integrate discreteness of the data in both the estimation of $m_0$ and the rejection of $p$-values.

\section*{Acknowledgements}
This work is part of project DO 2463/1-1, funded by the Deutsche Forschungsgemeinschaft. We  thank Etienne Roquain for insightful discussions and helpful suggestions which significantly improved the manuscript. We also thank Tobias Bedenk for valuable discussions on optimization problems and Florian Junge for  help with the simulations. We are also very grateful for the comments, questions and suggestions of two referees who gave us new ideas and made the paper much more thorough.

\section*{References}
\bibliographystyle{apalike}
\bibliography{biblio}

\appendix
\section{Auxiliary definitions, results {and proofs}}\label{appendix:auxres}
\subsection{Some definitions and results}
We recall some definitions and results of stochastic ordering following the presentation in  \cite{shaked2007stochastic}, to which we also refer the reader for further details.
We also recall a well-known bound on the inverse moment of the Binomial distribution.
\begin{definition}[Stochastic order]
	Let $X$ and $Y$ be two random variables such that
	\begin{align*}
		\P(X > x) \leq \P(Y > x) \quad \text{for all } x \in (-\infty, \infty),
	\end{align*}
	Then $X$ is said to be smaller than $Y$ in the usual stochastic order denoted by $X \stoorder Y$.
\end{definition}
	
An equivalent characterization of the stochastic order is that  $X \stoorder Y$ $\Leftrightarrow$ $\E[g(X)] \leq \E[g(Y)]$, for  all  non-decreasing functions $g : \R \rightarrow \R$ for which the expectations exist (see (1.A.7) in \cite{shaked2007stochastic})).	
%

\begin{definition}[Convex order] \label{app:def:cxorder}
	Let $X$ and $Y$ be two random variables such that 
		\begin{align*}
		\E(\phi( X)) \leq \E(\phi( Y)) \quad \text{for all convex functions } \phi : \R \rightarrow \R,
	\end{align*}
	provided the expectations exist. Then $X$ is said to be smaller than $Y$ in the convex order denoted as $X \cxorder Y$.
\end{definition}

The next results follows from the definition of convex ordering, see Chapter~3 of \cite{shaked2007stochastic}.

\begin{lemma}[Theorem 3.A.24 in \cite{shaked2007stochastic}] \label{app:lemma:SS:3-A-24}
	Let $X$ be a random variable with mean $\E X$. 
	Denote the left (right) endpoint of the support of $X$ by $l_X\left[u_X\right]$. 
	Let $Z$ be a random variable such that $\P\left\{Z=l_X\right\} = \left(u_X-\E X\right) /\left(u_X-l_X\right)$ and $\P\left\{Z=u_X\right\}=(\E X-$ $\left.l_X\right) /\left(u_X-l_X\right)$. 
	Then
	$$
		\E X \cxorder X \cxorder Z,
	$$
	where $\E X$ denotes a random variable that takes on the value $\E X$ with probability 1 (the left handside just restates Jensen's inequality).	
\end{lemma}

{The following Lemma is useful for proving an extremal property of the Bernoulli distribution w.r.t. $\cxorder$.}
{\begin{lemma}[Example 1.10.5 in \cite{MuellerStoyan2002}] \label{lemma:Example:Mueller:Stoyan}
	Let ${\mathcal M}_\mu^{[a,b]}$ be the class of all distributions on $[a,b]$ with mean $\mu$. Then this class contains a maximum element w.r.t. $\cxorder$ which is given by the two-point distribution 
	\begin{align*}
		F_{\textnormal{max}} = \frac{b-\mu}{b-a} \delta_{a} + \frac{\mu-a}{b-a} \delta_{b},
	\end{align*}
	where $\delta_{x}$ is the Dirac-measure in $x \in \R$.
\end{lemma}}

\begin{proposition}[Theorem 3.A.12 d) in \cite{shaked2007stochastic}] \label{app:lemma:SS:3-A-12}
	Let $X_1, X_2, \ldots, X_m$ be a set of independent random variables and let $Y_1, Y_2, \ldots, Y_m$ be another set of independent random variables. 
	If $X_i \cxorder Y_i$ for $i=1,2, \ldots, m$, then
	$$
	\sum_{j=1}^m X_j \cxorder \sum_{j=1}^m Y_j .
	$$
	That is, the convex order is closed under convolutions.
\end{proposition}

\begin{lemma} [Example 3.A.48 in \cite{shaked2007stochastic}] \label{app:lemma:SS:3-A-48}
	Let $X$ and $Y$ be Bernoulli random variables with parameters $p$ and $q$, respectively, with $0<p \leq q \leq 1$. 
	Then
	$$
	\frac{X}{p} \gecx \frac{Y}{q}.
	$$
\end{lemma}

\begin{lemma}[Inverse moment of the Binomial distribution]\label{prop:inverse:moments:parametric}
	For $n \in \N$, $s \in (0,1)$ we have
	\begin{align*}
		\E \left(\frac{1}{1+ \Bin(n,s)}\right) &= \frac{1-(1-s)^{n+1}}{(n+1)s}.
	\end{align*}
\end{lemma}

{\begin{proof}
		See e.g. \cite{benjamini2006adaptive}.
\end{proof}}

\begin{lemma} \label{lemma:Bernoulli:cx}
	For any $g \in \mathcal{G}$ we have $g(U) \cxorder \Bin (1,\nu)$ where $\nu = \E [g(U)]$ with $U \sim \unifrv[0,1]$.
\end{lemma}

\begin{proof}
		For $U \sim \unifrv[0,1]$ let  $\nu=\E [g(U)]$. Since $g([0,1])\subset [0,1]$ the distribution of $g(U)$ belongs to the class $\mathbb{M}_\nu^{[0,1]}$ of distributions on $[0,1]$ with expectation $\nu$. By Lemma \ref{lemma:Example:Mueller:Stoyan} this class possesses an extremal element w.r.t. $\cxorder$ which is given by the two-point distribution $F_{\textnormal{max}} = (1-\nu) \delta_{0} + \nu \delta_{1} \sim \Bin(1,\nu)$. Thus $g(U) \cxorder \Bin(1,\nu)$.
\end{proof}

\begin{remark}\label{prop:bias:variance:MSE:Storey}
	For Storey's estimator with $g_\lambda (t)=1_{(\lambda,1]}(t) $ we have
	\begin{align}
		\mse(\piZeroStorey(\lambda)) = \frac{1}{m(1-\lambda)^2} &  \left[ \frac{1}{m} - \pi_0 (1-\lambda)^2 +3(1-\pi_0) (1-F_1(\lambda))+\pi_0(1-\lambda) \right. \label{eq:MSE:Storey}\\
		& \left. \quad  +(m-1-m\pi_0)(1-\pi_0)(1-F_1(\lambda))^2  \vphantom{\frac{1}{m} } \right]. 		\notag
	\end{align}
\end{remark}

\begin{proof}
This follows immediately from Proposition \ref{prop:bias:variance:MSE} by taking $g_\lambda (t)=1_{(\lambda,1]}(t) $ in \eqref{eq:MSE:general}.
\end{proof}

\subsection{Proofs of results}

\begin{proof}[Proof of Proposition \ref{prop:plugin:control:general:g}]
	Since $\widehat{m}_0$ is coordinatewise non-decreasing, it is sufficient to verify \eqref{eq:IMC}.  
	For {any} $h \in \nullset$, monotonicity and super-uniformity give us $\widehat{m}_0( p_{0, h})  \gest 1/\nu+S_0$, where $\nu= min_{l \in \nullset \setminus \{ h \}} \nu_{l} $, and $S_0 = \sum_{\ell \in \nullset \setminus \{h\}} g_\ell(U_\ell)/\nu_\ell $ with $(U_\ell)_{\ell \in \nullset}$ i.i.d random variables distributed according to $\unifrv[0,1]$. 
	By Lemma \ref{lemma:Bernoulli:cx} we have $g_\ell(U_\ell) \cxorder \Bin (1,\nu_\ell)$ and Lemma~\ref{app:lemma:SS:3-A-48} gives $\Bin(1,\nu_i)/\nu_i \cxorder \Bin(1,\nu)/\nu$. 
	Since the convex ordering is preserved under convolutions (see Lemma~\ref{app:lemma:SS:3-A-12}) we obtain $ \nu S_0 \cxorder \Bin (m_0-1, \nu)$. 
	Finally, the mapping $x \mapsto \nu/(1+x)$  is convex on $[0, \infty)$ and therefore from the Definition~\ref{app:def:cxorder} of $\cxorder$ we obtain that 
	\begin{align} \label{eq:proofprop3.1}
		\E \left( \frac{1}{\widehat{m}_0( p_{0, h})}\right) &\le \E \left( \frac{1}{\frac{1}{\nu}+S_0}\right) = \E \left( \frac{\nu}{1+\nu S_0}\right) \le \E \left( \frac{\nu}{1+\Bin (m_0-1, \nu)}\right) \le \frac{1}{m_0},
	\end{align}
	where the last bound is a well-known result for the inverse moment of Binomial distributions (see Lemma \ref{prop:inverse:moments:parametric}) so that \eqref{eq:IMC} is proved. 	The statement on plug-in FDR control now follows from Theorem \ref{thm:IMC}.
\end{proof}

{By taking $g(u)=\ind{u>\lambda}$ we have $\nu S_0 \sim \Bin (m_0-1, \nu)$ and therefore the second inequality from the right  in \eqref{eq:proofprop3.1} can be replaced by an equality. Thus, it may be tempting to conclude that $\mStorey$ is optimal. In the case of a Dirac-Uniform constellation of $p$-values (see \cite{BR2009})  this is indeed true, since $\nu \mStorey \sim 1 + \Bin (m_0-1, \nu)$ and therefore the left inequality in \eqref{eq:proofprop3.1} can also be replaced by an equality. In more general settings however,  other choices of $g$ may be better. 

\begin{proof}[Proof of proposition \ref{prop:plugin:discrete:control:general:g}]
	In the first step we show that for any $i \in \nullset$ we can find a function $g_i \in \mathcal{G}$ such that
		\begin{align}
			g(p_i) &\sim g_i(U)  \qquad \text{and} \qquad \nu^{\text{resc}}_i = \E g_i(U), \label{eq:proof:discrete:control:general:g:1}
		\end{align}
		where $U \sim \unifrv[0, 1]$. This will allow us to replicate the proof of Proposition \ref{prop:plugin:control:general:g} for establishing \eqref{eq:IMC} (see second step below).   To this end we define  $g_i : [0,1] \rightarrow [0,1]$ by $g_i(y) = g \circ F^{-1}_i (y)$ for $y \in (0, 1]$, where $F^{-1}_i (y) = \inf \{x \in \R: F_i(x) \ge y \}$ is the generalized inverse of $F_{i}$, and set $g_i(0)=g(0)$. 	Since $g \in \mathcal{G}$ and $F^{-1}_i $ are both nondecreasing, so is $g_i$.
	For $i \in \mathcal{H}_{0}$ we have  $p_{i} \sim F^{-1}_i (U)$ {by Proposition 2 in \cite{EmbrechtsHofert2013}}, so that $g_{i}(U) \sim g(p_{i})$ which implies $ \E [g_{i}(U)] = \E_{p_i \sim F_i} [g(p_i)] = \nu^{\text{resc}}_i $, i.e \eqref{eq:proof:discrete:control:general:g:1} holds true.
	
	In the second step we show that $\widehat{m}^{\text{resc}}_0$ satisfies  \eqref{eq:IMC}, thereby guaranteeing plug-in FDR control. We have
	\begin{align}
		\widehat{m}^{\text{resc}}_0 (p_{0,h}) &\ge
		\frac{1}{min(\nu^{\text{resc}}_1, \ldots, \nu^{\text{resc}}_m ) } + \sum_{i \in \nullset \setminus \{h\}} \frac{g(p_i)}{\nu^{\text{resc}}_i}
		\gest
		 \frac{1}{min_{i \in \nullset \setminus \{ h \}} \nu^{\text{resc}}_{i} } + \sum_{i \in \nullset \setminus \{h\}} \frac{g_i(U_i)}{\nu^{\text{resc}}_i}, \label{eq:proof:discrete:control:general:g:2}
	\end{align}
	where we have used the representation \eqref{eq:proof:discrete:control:general:g:1} from the first step with independent $U_i$'s. The r.h.s. of \eqref{eq:proof:discrete:control:general:g:2} is immediately seen to be identical with the random variable $1/\nu+S_0$ in the proof of Proposition \ref{prop:plugin:control:general:g}. From this  we obtain  \eqref{eq:proofprop3.1} which implies \eqref{eq:IMC} and thus the statement on plug-in FDR control follows.
	
	The claim that $\mresc \le \widehat{m}_0$ follows since $\nu_1, \ldots, \nu_m \ge \nu (g)$ by super-uniformity and non-decreasingness of $g$.
\end{proof}

\begin{lemma}[Properties of $\bar{g}$] \label{lemma:prop:g:bar}
Let $p \sim F$ be a discrete super-uniform distribution on $[0,1]$ with finite support $\mathcal{S}=\{0=s_0<s_1 < s_2 < \ldots < s_L=1\}$. Let the function $g \in \mathcal{G}$ and  define $\bar{g}_{F} $ as in \eqref{def:g:bar:F}. Then 
\begin{itemize}
	\item[a)] $\bar{g}_{F} \in \mathcal{G}$,
	\item[b)] $\bar{g}_{F}(p)  \lecx g(U)$, where $U\sim \unifrv(0,1)$,
	\item[c)] $\bar{g}_{F}(s_k) \le g(s_k)$ for $k=1, \ldots, L$,
	\item[d)] $\E \bar{g}_{F}(p) = \E g(U)$.
\end{itemize}
\end{lemma}

\begin{proof}[Proof of Lemma \ref{lemma:prop:g:bar}]
To verify the non-decreasingness of $\bar{g}_{F} $, we observe that 
\begin{align*}
	\bar{g}_{F}(t)&= \sum_{k=1}^{L} a_k \cdot \ind{(s_{k-1},s_k]}(t) \qquad 	\text{with} \qquad 	a_k = \frac{1}{F(s_k)-F(s_{k-1})} \int_{F(s_{k-1})}^{F(s_k)} g(s) ds
	\intertext{and therefore $a_k \in [0,1]$ and for $u,v \in \{1, \ldots,L\}$ with $u<v$ we have }
	a_u & \le \sup_{s \in (F(s_{u-1}),F(s_u)]} g(s) \le \inf_{s \in (F(s_{v-1}),F(v_u)]} g(s) \le a_v.
\end{align*}
To prove  statement b) we define for $u \in (0,1]$ 
\begin{align*}
	I(u) &= \{k \in \{1, \ldots,L\} | u \in  (F(s_{k-1}),F(s_k)] \}.
\end{align*}
and set $I(0)=0$. Thus, for $U \sim \unifrv[0,1]$ we have $p \sim s_{I(U)}$. From the above expression of $\bar{g}_{F}$  we have for all $k \in \{1, \ldots,L\} $
\begin{align*}
	\bar{g}_{F}(p) | I(U)=k &= a_k = \frac{1}{F(s_k)-F(s_{k-1})} \int_{F(s_{k-1})}^{F(s_k)} g(s) ds = \E (g(U)|I(U)=k)
	\intertext{which implies, by \cite{shaked2007stochastic}[Theorem 3.A.24] that}
	\bar{g}_{F}(p) | I(U)=k & \lecx g(U)|I(U)=k.
\end{align*} 
The claim now follows since the convex order is closed under mixtures (see \cite{shaked2007stochastic}[Theorem 3.A.12]).

Statement c) follows from $\bar{g}_{F}(0) =0$ and from the fact that for all $s \in (F(s_{k-1}),F(s_k)]$ we have $g(s)\le g(F(s_k)) \le g(s_k)$ because $F$ is super-uniform and $g$ is non-decreasing. Therefore, $\bar{g}_{F}(s_k) =a_k\le g(s_k)$.

Statement d) follows from statement b) (see (3.A.2) in \cite{shaked2007stochastic}).
\end{proof}

\begin{proof}[Proof of Proposition \ref{prop:plugin:discrete:control:mean:g}	]
	By statement a) of  Lemma \ref{lemma:prop:g:bar} each $\bar{g}_{F_i}$ is non-decreasing, implying that $\mmean$  is coordinatewise non-decreasing, so that it is sufficient to verify \eqref{eq:IMC}. 	For {any} $h \in \nullset$, we have  
	\begin{align*}
		\mmean ( p_{0, h})  &\gest \frac{1}{\nu (g)} \left(1+ \sum_{\ell \in \nullset \setminus \{h\}} \bar{g}_{F_\ell}(p_\ell) \right)
		\intertext{and therefore}
		\E \left(\frac{1}{\mmean ( p_{0, h})}\right)  &\le \nu (g) \cdot \E \left( \frac{1}{1+ \sum_{\ell \in \nullset \setminus \{h\}} \bar{g}_{F_\ell}(p_\ell)}\right) \le \nu (g) \cdot \E \left( \frac{1}{1+ \sum_{\ell \in \nullset \setminus \{h\}} g(U_\ell)}\right).
	\end{align*}
The last inequality can be verified as follows. Let $(U_\ell)_{\ell \in \nullset}$ be i.i.d random variables distributed according to $\unifrv[0,1]$. Statement b) of \ref{lemma:prop:g:bar} gives us $\bar{g}_{F_\ell}(p_\ell)  \lecx g(U_\ell)$ for all $\ell \in \nullset $ and since the convex order is closed under convolutions by Proposition  \ref{app:lemma:SS:3-A-12}, this implies that $\sum_{\ell \in \nullset \setminus \{h\}} \bar{g}_{F_\ell}(p_\ell) \lecx \sum_{\ell \in \nullset \setminus \{h\}} g(U_\ell)$. The last inequality now follows from the convexity of the mapping $x \mapsto \frac{1}{1+x}$. The remainder of the proof of FDR control is identical to the proof of Proposition \ref{prop:plugin:control:general:g} with $g_1= \ldots=g_m=g$.
	
	The claim that $\mmean\le \widehat{m}_0 $ follows directly from statement c) of Lemma \ref{lemma:prop:g:bar}.
\end{proof}

\begin{proof}[Proof of Proposition \ref{coro:randomized:plugin:control}]
	The proof of (a) uses Theorem \ref{thm:IMC}. First, we show that $\mrand (p_1, \ldots, p_m)$ is coordinatewise non-decreasing. 
	For fixed $(u_1, \ldots, u_m) \in [0,1]^m$ each (realized) randomized $p$-value $r_i=r_i(p_i,u_i)$ is non-decreasing in $p_i$. Since $\widehat{m}_0 \in \mathcal{F}$ is coordinatewise non-decreasing in $(p_1, \ldots, p_m)$, the function $1/\widehat{m}_0(r_1(\cdot,u_1), \ldots, r_m(\cdot,u_m))$ is coordinatewise decreasing for all $(u_1, \ldots, u_m) \in [0,1]^m$ and so is its expectation which implies that $\mrand$ is coordinatewise non-decreasing.
	To establish \eqref{eq:IMC},   we denote for $h \in \nullset$ by $r_{0,h}$ the set of randomized $p$-values $(r_1, \ldots,r_m)$, where $r_h$ has been replaced by $0$. 
	By the definition of $\mrand$ we have
	\begin{align*}
		\E_{(p_1, \ldots, p_m)} \left[\frac{1}{\mrand(p_{0,h})}\right] &= \E_{(p_1, \ldots, p_m)} \left[ \E_{(U_1, \ldots, U_m)} \frac{1}{\widehat{m}_0(r_{0,h})} \right]=\E_{(r_1, \ldots, r_m)} \left[\frac{1}{\widehat{m}_0(r_{0,h})} \right].
	\end{align*}
	where the second equality follows from the fact that for super-uniform $p$-value $p_{h} = 0$,  the associated randomized $p$-value $r(p_{h}, u) = 0$ (a.s) by Definition~\eqref{eq:def:randomp}.
	Since the $(r_1, \ldots, r_m)$ are mutually independent and uniform under the null and $\widehat{m}_0 \in \mathcal{F}$, the bound \eqref{eq:IMC} for $\widehat{m}_0(r_{0,h})$ now follows 
	since $\widehat{m}_0$ satisfies the conditions of Theorem \ref{thm:IMC}. Therefore, the r.h.s. of the last equation can be bounded by $1/m_0$ and {plug-in FDR control} for $\mrand$ now follows from Theorem \ref{thm:IMC}. 
	
	{To see that  statement {(b)} holds true, observe that since $\widehat{m}_0$ is coordinatewise non-decreasing and $r_i(p_i, U_i) \le r_i(p_i,0)=p_i$ we have $\widehat{m}_0 (r_1(p_1,U_1), \ldots, r_m( p_m,U_m)) \le \widehat{m}_0(p_1,\ldots,p_m)$ and therefore the r.h.s. of \eqref{def:m0:rand} is bounded by $\widehat{m}_0(p_1,\ldots,p_m)$ $(a.s.)$.}
	
	{For the proof of statement (c) first consider a single discrete $p$-value $p \sim F$  on $[0,1]$ with support  $\mathcal{S}=\{0=s_0<s_1 < s_2 < \ldots < s_L=1\}$ and assume (as usually) that for $x \in \mathcal{S}$   we have $F(x)=x$. Suppose $s_k$ is the observed value of $p$. Then we have $r(s_k,U) \sim U(s_{k-1},s_{k})$ and therefore 
	\begin{align}
		\E_U& g(r(s_k,U)) = \E g (U(s_{k-1},s_{k}))= \bar{g}_{F}(s_k) \label{eq:proof:b}
\end{align}
by \eqref{def:g:bar:F}. To prove statement (c), we show first that for $\widehat{m}_0 \in \mathcal{F}_0$ as in \eqref{eq:def:class:estimators:0} with  $g \in \mathcal{G}$ we have 
\begin{align}
\E_{U} \widehat{m}_0 (r(p,U)) &=\mmean (p_1, \ldots, p_m). \label{eq:identity:EU:mean}
\end{align}
This follows from
\begin{align*}
	\E_{U} \widehat{m}_0 (r(p,U)) &=  \frac{1}{\nu(g)} \left(1+ \sum_{i=1}^m \E_{U_i}g (r_i(p_i,U_i))  \right)=\frac{1}{\nu(g)} \left(1+ \sum_{i=1}^m \bar{g}_{F_i}(p_i) \right) = \mmean (p_1, \ldots, p_m),
\end{align*}
where the second equality is due to \eqref{eq:proof:b} and the last  inequality is the definition of $\mmean$ in \eqref{eq:def:discrete:adjusted:estimator}. Jensen's inequality gives us $\mrand  (p_1, \ldots, p_m)\le \E_{U} \widehat{m}_0 (r(p,U))$ which completes the proof.}
\end{proof}

The following result and its proof are due to Tobias Bedenk (personal communication, March 2026).
\begin{proposition}\label{prop:Tobias}
\begin{itemize}
	\item[a)] The optimization program \eqref{eq:optimisation:bias:discrete:general} can be transformed into the following Quadratically Constrained Linear Program.
	\begin{mini}
		{\substack{\alpha, \beta_1, \ldots, \beta_m\\ \bw_1, \ldots, \bw_m}}{\alpha + (1-\pi_0) \sum_{i=1}^m \beta_i,}{}{}
		\addConstraint{\alpha \cdot \bw_i^Tx_i }{\geq 1 \quad \forall i=1,\dots,m  }
		\addConstraint{\bw_i^T (\beta_ix_i - y_i)}{ = 0 \quad \forall i=1,\dots,m}
		\addConstraint{\|\bw_i\|_1}{\leq 1 \quad \forall i=1,\dots,m}
		\addConstraint{\bw_i}{\geq 0 \quad \forall i=1,\dots,m}\label{eq:OP:bias:equivalent} 
	\end{mini}
	\item[b)] The program \eqref{eq:OP:bias:equivalent}  is NP-hard.
\end{itemize}	
\end{proposition}

\begin{proof} For statement a),	let  $w_1, \ldots , w_m $ be a solution of  \eqref{eq:optimisation:bias:discrete:general}. Now define 
\[\alpha := \frac{1}{\min\{\bw_1^Tx_1, \dots \bw_m^Tx_m\}}, \quad \text{and} \qquad \beta_i := \frac{\bw_i^Ty_i}{\bw_i^Tx_i},\]
noting that $\alpha$ and the $\beta_i$'s can only be positive. With the above definition, $\alpha$, the $\beta_i$'s and the $\bw_i$'s  are a solution of the following program:
\begin{mini}
	{\substack{\alpha, \beta_1, \ldots, \beta_m\\ \bw_1, \ldots, \bw_m}}{\alpha + (1-\pi_0) \sum_{i=1}^m \beta_i ,}{}{}
	\addConstraint{\alpha}{ = \frac{1}{\min\{w_1^Tx_1, \dots w_m^Tx_m\}}}{}
	\addConstraint{\beta_i}{= \frac{\bw_i^Ty_i}{\bw_i^Tx_i}}{\quad \quad \forall i=1,\dots,m}
	\addConstraint{\|\bw_i\|_1}{\leq 1}{\quad \forall i=1,\dots,m}
	\addConstraint{\bw_i}{\geq 0}{\quad \forall i=1,\dots,m}\label{eq:OP:bias:equivalent:proof}
\end{mini}
We will now transform the first and the second constraints  in order to obtain quadratic constraints as in \eqref{eq:OP:bias:equivalent}. 
Since $\alpha =\max\{\frac{1}{w_1^Tx_1}, \dots, \frac{1}{w_m^Tx_m}\}$ it is clear that
%
%
\begin{align*}
	\alpha &\geq \frac{1}{\bw_i^Tx_i} \quad \forall i=1,\dots,m
	\intertext{needs to hold which is equivalent to} 
	\alpha \cdot \bw_i^Tx_i &\geq 1 \quad \forall i=1,\dots,m.
\end{align*}
Since we have, for each $\beta_i$ that
\begin{align*}
	\ &\beta_i = \frac{\bw_i^Ty_i}{\bw_i^Tx_i} \quad 
	\Leftrightarrow \quad \beta_i \cdot \bw_i^Tx_i = \bw_i^Ty_i \quad 	\Leftrightarrow \quad \bw_i^T(\beta_ix_i - y_i) = 0
\end{align*}
we can re-express program \eqref{eq:OP:bias:equivalent:proof} as \eqref{eq:OP:bias:equivalent} and the values of the $\bw_i$'s of an optimal solution of \eqref{eq:OP:bias:equivalent} are also optimal for   the original program \eqref{eq:optimisation:bias:discrete:general}. 

For statement b), consider the first constraint of the program \eqref{eq:OP:bias:equivalent}, which can be rewritten as 
\[\alpha \bw_i^Tx_i = \frac{1}{2} (\alpha, \bw_i^T) \left(\begin{array}{cc}
	0 & x_i^T \\
	x_i & 0
\end{array} \right) \left( {\alpha \atop \bw_i} \right) \geq 1\]
The matrix in this inequation is of rank 2 and hence has the eigenvalue 0 with multiplicity $n-2$. The other two eigenvalues are $\pm \|x_i\|$. Since one eigenvalue is positive, and the other negative, the matrix is neither positive nor negative semidefinite, i.e. the quadratic function defining the constraint is not convex and not concave.  The second constraint can be transformed analoguously. In general, programs with neither positive nor negative semindefinite constraints  are  NP-hard, see e.g. \cite{Burer2012}.
\end{proof}

\begin{proof}[Proof of Proposition \ref{prop:bias:variance:MSE}]
	Since 
	\begin{align*}
		\widehat{\pi}_0 &=\frac{1}{m\nu}+\frac{1}{m} \sum_{i \in \nullset} \frac{g(p_i)}{\nu}+\frac{1}{m} \sum_{i \in \altset} \frac{g(p_i)}{\nu},
		\intertext{we have}
		\E \widehat{\pi}_0 &=\frac{1}{m\nu}+\frac{1}{m} |\nullset|+\frac{1}{m} |\altset|\frac{\E_1 g(p_i)}{\nu} = \frac{1}{m\nu} + \pi_0+ (1-\pi_0)\frac{\E_1 g(p_i)}{\nu},
		\intertext{and the statement for $\bias(\widehat{\pi}_0)$ follows immediately. For the variance, independence of the $p$-values  yields }
		\var (\widehat{\pi}_0) &= \frac{1}{m^2 \nu^2} \left(|\nullset| \var_0 g(p) + |\altset| \var_1 g(p)\right) = \frac{1}{m \nu^2} \left(\pi_{0} \var_0 g(p) +(1-\pi_{0}) \var_1 g(p)\right),
	\end{align*}
	and the formula for the $\mse (\widehat{\pi}_0)$ follows by the bias-variance decomposition and some basic algebra.
\end{proof}

For the proof of Proposition \ref{prop:MSE:quadratic:oracle} we first prove the following Lemma.

\begin{lemma} \label{prop:MSE:quadratic:oracle:appendix}
	Assume a two-groups model as in Proposition	\ref{prop:MSE:quadratic:oracle} and let $K \in \N$ and $\blambda \in \Lambda$ be fixed.	Then we have, for any $g=g_{\bw,\blambda} \in {\mathcal G} (\blambda)$ with $\bw \in {\mathcal W} $: 
		\begin{align}
			\E_1g(p) &= \sum_{k} w_k(1-F_1(w_k))= \bw^T \cdot \eta,\label{eq:prop:MSE:quadratic:oracle:appendix:1}\\
			[\E_1 g(p)]^2 &= \sum_{k,\ell}w_k(1-F_1(\lambda_k)) (1-F_1(\lambda_{\ell})) w_{\ell}= \bw^T \cdot Y \cdot \bw,\label{eq:prop:MSE:quadratic:oracle:appendix:2}\\
			\E_0[g(p)^2] &= \sum_{k,\ell}w_k(1-\max(\lambda_k,\lambda_{\ell})) w_{\ell} = \bw^T \cdot X^0 \cdot \bw,\label{eq:prop:MSE:quadratic:oracle:appendix:3}
			\intertext{and}
			\E_1[g(p)^2] &=  \sum_{k,\ell}w_k(1-F_1(\max(\lambda_k,\lambda_{\ell}))) w_{\ell}= \bw^T \cdot X^1 \cdot \bw,\label{eq:prop:MSE:quadratic:oracle:appendix:4}
			\intertext{where the vector $\eta \in [0,1]^K$ and matrices $X^0,X^1,Y \in [0,1]^{K \times K}$ are defined by }
			\eta_k &= 1- F_1(\lambda_k),\notag\\
			X^0_{k, \ell} &= 1-\max(\lambda_k,\lambda_{\ell}),\notag\\
			X^1_{k, \ell} &=1-F_1(\max(\lambda_k,\lambda_{\ell})),\notag
			\intertext{and}
			Y_{k, \ell} &=(1-F_1(\lambda_k))\cdot (1-F_1(\lambda_{\ell})).\notag
		\end{align}
\end{lemma}

\begin{proof}
Since
\begin{align*}
g_{\bw,\blambda}(t) &= \sum_{k=1}^K w_k \cdot 1_{(\lambda_k,1]}(t) ,
\intertext{\eqref{eq:prop:MSE:quadratic:oracle:appendix:1} follows from the linearity of expectation and implies  \eqref{eq:prop:MSE:quadratic:oracle:appendix:2}. Since}
g_{\bw,\blambda}(t)^2 &= \sum_{k,\ell}w_k \cdot 1_{(\lambda_k,1]}(t) \cdot 1_{(\lambda_\ell,1]}(t)\cdot  w_{\ell} = \sum_{k,\ell}w_k \cdot 1_{(\max(\lambda_k,\lambda_\ell),1]}(t) \cdot  w_{\ell},
\intertext{we have, for $a\in \{0,1\}$}
\E_a [g_{\bw,\blambda}(t)^2] &= \sum_{k,\ell}w_k \cdot (1-F_a(\max(\lambda_k,\lambda_\ell)) \cdot   w_{\ell},
\end{align*}
which yields \eqref{eq:prop:MSE:quadratic:oracle:appendix:3} and \eqref{eq:prop:MSE:quadratic:oracle:appendix:4}.
\end{proof}

\begin{proof}[Proof of Proposition \ref{prop:MSE:quadratic:oracle}] Let $\eta \in [0,1]^K$ and  $X^0,X^1,Y \in [0,1]^{K \times K}$ be defined as in Lemma \ref{prop:MSE:quadratic:oracle:appendix} and  define a vector $d \in \R^K$ and a matrix $C \in \R^{K \times K}$ by 

\begin{align*}
d &= 2(1-\pi_0)\eta,\\
C &= 2\left[(m-1-m\pi_0)(1-\pi_0)Y
+\pi_0X^0
+(1-\pi_0)X^1\right].
\end{align*}

Then $C$ is symmetric and, for every $\bw\in\mathcal W$, Proposition~\ref{prop:bias:variance:MSE} gives
\begin{align*}
m\nu^2\,\mse(\piZeroConv)
&=
\frac1m-\pi_0\nu^2
+2(1-\pi_0)\E_1 g(p)
\\
&\qquad
+(m-1-m\pi_0)(1-\pi_0)(\E_1g(p))^2
+\pi_0\E_0g(p)^2
+(1-\pi_0)\E_1g(p)^2.
\end{align*}

Applying Lemma~\ref{prop:MSE:quadratic:oracle:appendix},
\begin{align}
m\nu^2\,\mse(\piZeroConv)
&=
\frac1m-\pi_0\nu^2
+2(1-\pi_0)\bw^T\eta
\notag \\
&\qquad
+(m-1-m\pi_0)(1-\pi_0)\bw^TY\bw
+\pi_0\bw^TX^0\bw
+(1-\pi_0)\bw^TX^1\bw
\notag \\
&=
\frac1m-\pi_0\nu^2
+\bw^Td
+\frac12\bw^TC\bw, \label{eq:mse:two:groups:matrix:appendix}
\end{align}
where $\nu=\nu_{\bw,\blambda}= \bw^T \cdot (1- \blambda)$.
\end{proof}

\section{Some more real data analyses } \label{appendix:sec:additional:analyses:real:data} 
 In this appendix we present some more data analyses similar to those performed on  the IMPC data in Section \ref{sec:real:data:analysis}. All data-sets used here are contained in the R package \texttt{DiscreteDatasets} (see \cite{DiscreteDatasets2024}) which also contains more detailed descriptions. For a brief summary  we just  give the  following background information: 
\begin{itemize}
	\item The \texttt{amnesia} data-set, provided by the UK Regulatory Agency, includes adverse drug reactions due to medicines and healthcare products. It contains the number of reported cases of amnesia as well as the total number of adverse events reported for each of the $m = 2446$ drugs in the database.   \cite{DDR2018} investigated the association between reports of amnesia and suspected drugs by performing for each drug (two-sided) FETs.
	\item The \texttt{listerdata} data-set contains data from around 22,000 cytosines, each of which is under two conditions. For each cytosine under each condition, there is only one replicate. The discrete count for each replicate can be modeled by binomial distribution, and Fisher's exact test can be applied to assess if a cytosine is differentially methylated. The filtered data lister contains cytosines whose total counts for both lines are greater than 5 and whose count for each line does not exceed 25.	
	\item The \texttt{hiv} data-set contains data  for $m=118$ mutated amino acids in 73 subjects with two different types of HIV. For each mutated amino acid the data is represented by a   $2 \times 2$ table, which is analyzed by a (two-sided) FET.
	\item The \texttt{airway} data-set contains read counts of $m=63677$ genes for airway smooth muscle cell-lines RNA-Seq experiment under two treatment conditions. The result for each specific gene is summarized by a $2 \times 2$ table, which is analyzed by a two-sided FET.	
\end{itemize}

\begin{center}
	\begin{figure}[htbp!]  
		\includegraphics[width=.9\linewidth]{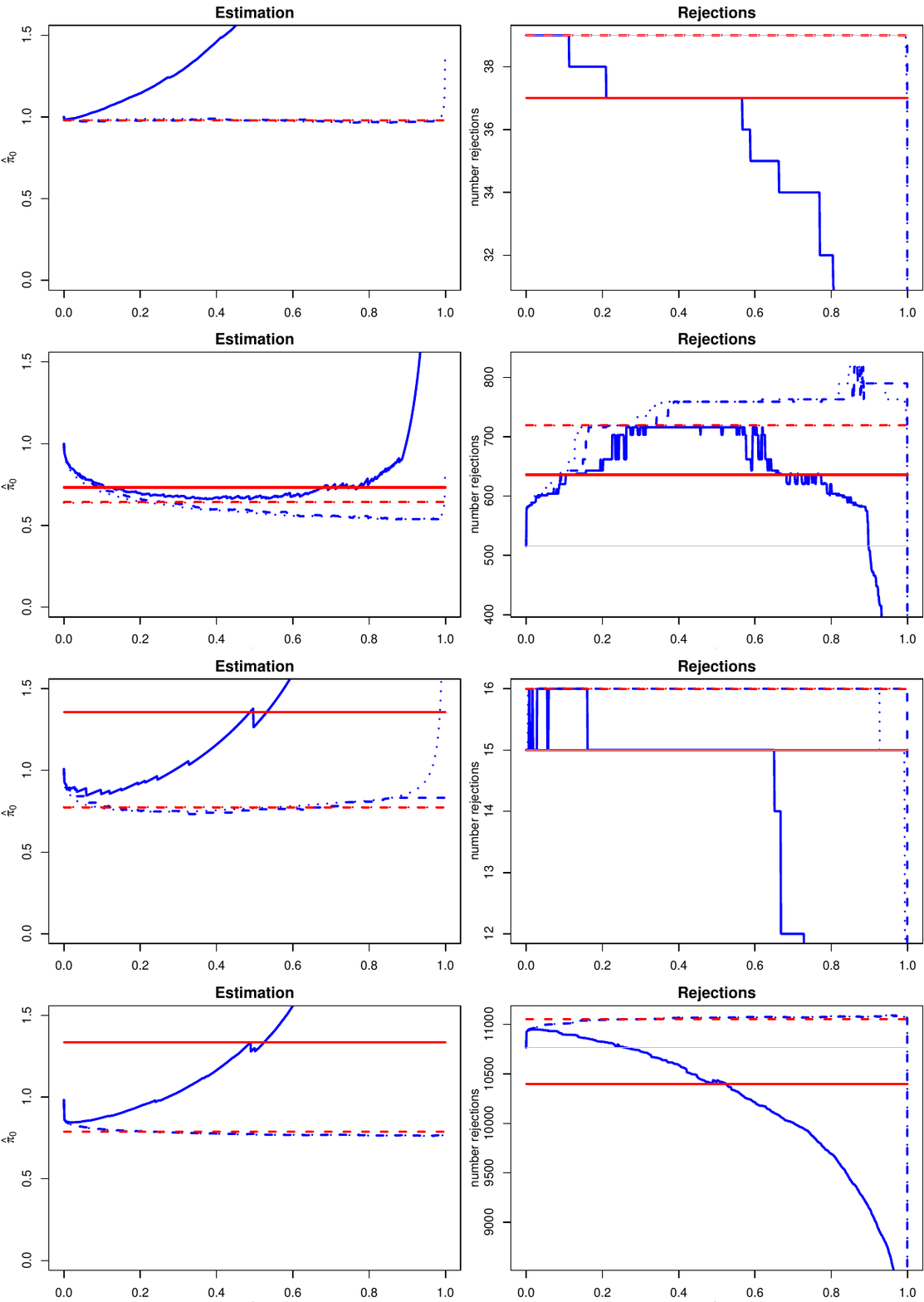}
		\caption{$\pi_{0}$-estimators (left panels) and number of discoveries  for the plug-in BH procedure (right panels) for the  amnesia,  lister, hiv, and airway data-sets (from top to bottom). Blue lines represent versions of the Storey estimator, red lines represent versions of the PC estimator. For both estimators the solid lines represent the base (continuous) estimator, the dashed lines represent the rescaled \eqref{eq:def:discrete:rescaled:estimator} and the dotted lines the conditional mean \eqref{eq:def:discrete:adjusted:estimator} modifications for discrete data. The thin grey line in the right panel represents the number of discoveries by the (non-adaptive) BH procedure. The results for the randomized estimator \eqref{def:m0:rand} are practically identical with those for the conditional mean estimator  and have therefore been omitted from the graphs for better readability.}
		\label{fig:MultipleDiscretePlots}
	\end{figure}
\end{center}

In our analysis we have not included the estimator of \cite{chen2018multiple}. The reason for this is that this estimator is guaranteed to be conservative only under 	a very restrictive condition: the so-called 'guiding values' $\tau_j$ must lie in the intersection
	of the supports of all $p$-value distributions. This condition was introduced in a correction by \cite{ChenDoerge2020} after  \cite{BiswasLetter2020} pointed out an error in the proof of plug-in FDR control. 	We verified whether this condition holds in the real data-sets used above  and found that it is violated in all cases. More specifially, the condition fails as soon as two $p$-value supports intersect only at the point 1, which renders the estimator of \cite{chen2018multiple, ChenDoerge2020} inapplicable. This situation occurs in all our data-sets: For
	example, in the amnesia data-set, $p$-values 19 and 24 share only the value 1 in their
	supports; similar cases arise for $p$-values 35 and 37 in the lister dataset, $73$ and 78 in the
	HIV dataset, 192 and 196 in the airway dataset, and 35 and 36 in the IMPC dataset.
	
Figure \ref{fig:MultipleDiscretePlots} illustrates the results of our analyses. By and large, we can draw conclusions similar to the ones for the IMPC data in Section \ref{sec:real:data:analysis}. More specifically, we can see that taking discreteness into account is beneficial both for estimation and inference purposes and the gains in efficiency can in some cases be large, irrespective of which specific adjustment method was chosen. For Storey's estimator, the discrete adjustments lead to more stable behavior over the range of tuning parameters. 

Whereas the performance of the base Pounds-Cheng estimator seems  quite conservative when compared to the base Storey estimator, its discrete variants  perform quite well, yielding robust yet comparable results to the discrete Storey estimators, at least on the data sets we have analyzed.

\section{More details on the Pounds-Cheng estimator} \label{appendix:ssec:ComparePCNew:PCZZD}
	{\cite{PC2006} introduced their estimator \eqref{def:m0:PC:2006} primarily to obtain a robust estimate of FDR.} 
	To the best of our knowledge, the only previously available result on plug-in FDR control was obtained by \cite{ZZD2011}, who defined the following  modified version of $\mPCOrig$:
	\begin{align}
		\label{def:PCZZD}
		\mZZKB& = C(m) \cdot \min \left[ m, \max \left(s(m), 2 \cdot \sum_{i =1}^m p_i \right) \right], 
			\end{align}
	where the correction factors $C(m)$ and $s(m)$ are chosen in such a way  that \eqref{eq:IMC} holds. 
	{However, determining} these factors is non-trivial and requires extensive use of numerical integration and approximations methods (see Supplement B in \cite{ZZD2011} for further details) so that no simple representation of $C(m)$ and $s(m)$ is available (for selected values of $m$, Table S1 in \cite{ZZD2011} lists values for the correction factors).} 
Here we present some numerical results, comparing the performance of $\mPCNew$ \eqref{eq:def:PC:new} and $\mZZKB$ \eqref{def:PCZZD} for $m=500$, where the correction factors $C(500)=1.011709$ and $s(500)=98$ have been taken from Table S1 in \cite{ZZD2011}. 

We first analyze the two estimators on simulated data in a one-sided Gaussian testing setting where we observe realizations of independent rv's $X_{1}, \dots, X_{m_0} \sim N(0,1)$ and $X_{m_0 +1}, \dots, X_{500} \sim N(1.5,1)$ for $1000$ Monte-Carlo simulation runs and a varying range of $m_0 =50,100, \dots, 450$. 
We obtain $500$ $p$-values by testing the null hypotheses $H_{0, i} : \mu = 0$ vs. the alternatives $H_{1, i} : \mu  > 0$ simultaneously for all $i \in \{1, \dots, 500\}$ and calculate $\mPCNew$ and $\mZZKB$ as well as the number of rejections obtained from the plug-in BH procedure in \eqref{eq:khat:BH} with $\alpha=0.05$. 

\begin{center}
	\begin{figure}[hbt]
		\center
		\begin{tabular}{cc}
			\includegraphics[scale=0.12]{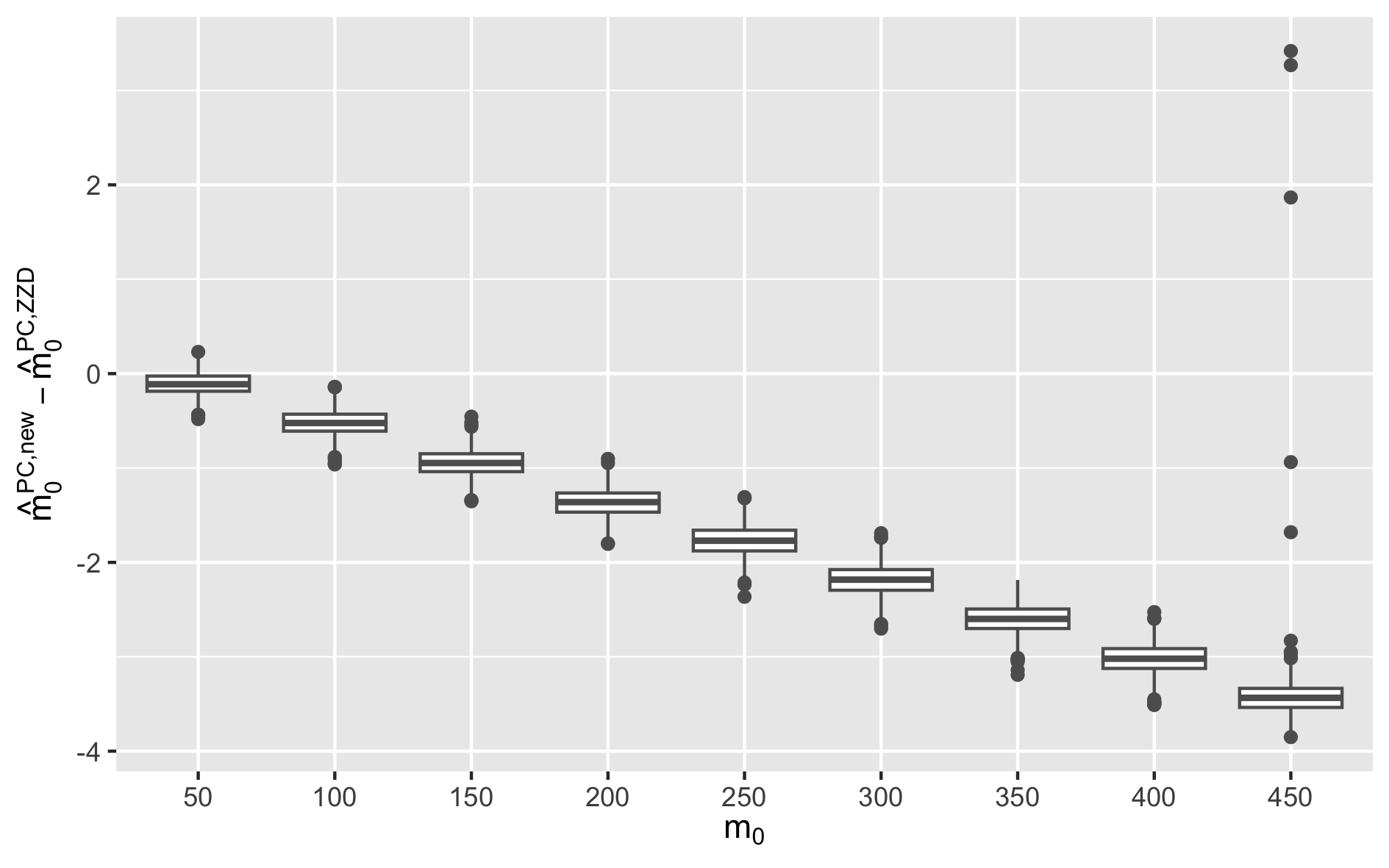} & \includegraphics[scale=0.12]{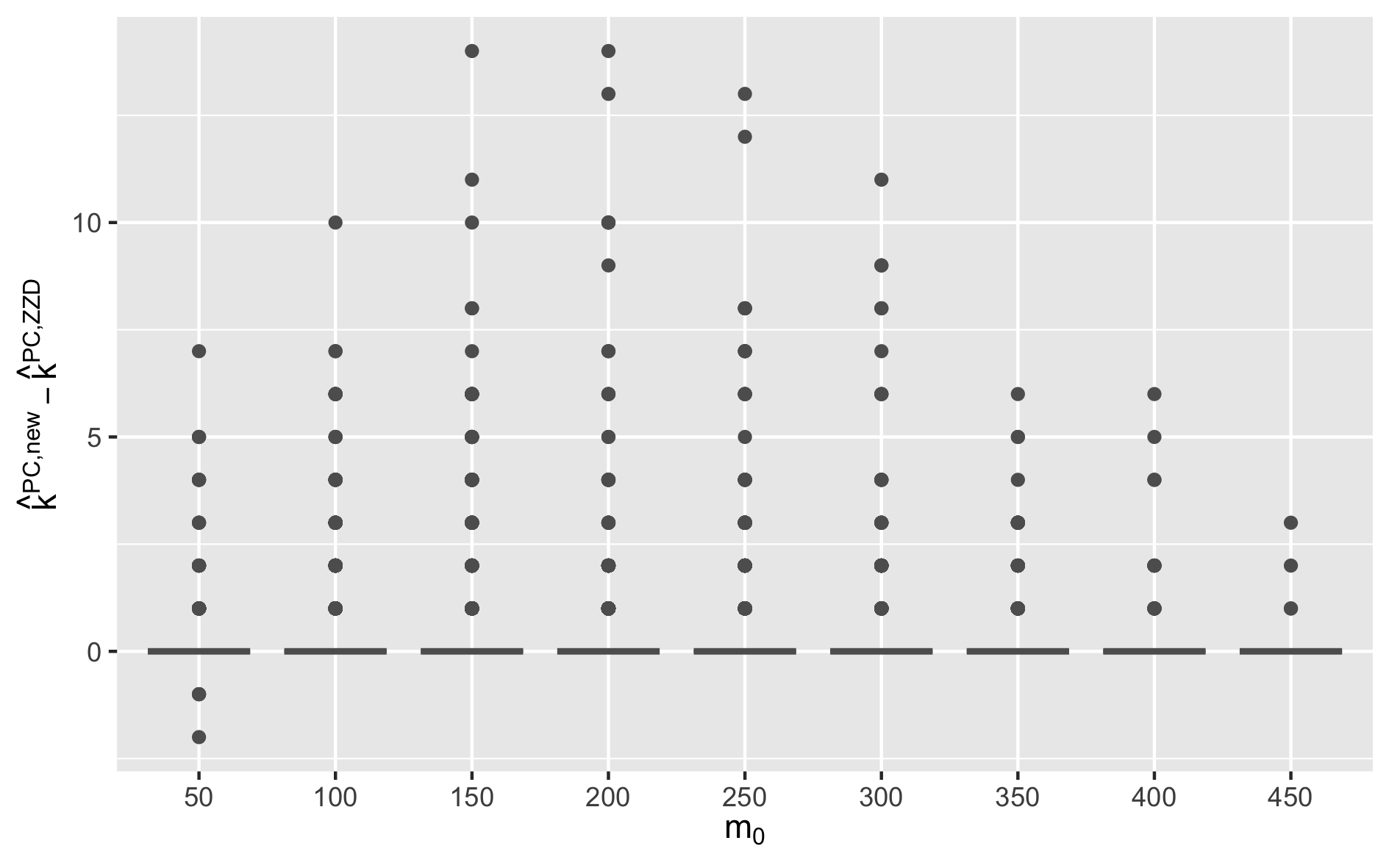}
		\end{tabular}
		\caption{Box plots for the difference between $\mPCNew$ and $\mZZKB$ for point estimation (left) and rejection numbers for the plug-in BH procedures (right) against a range of true $m_0 =50,100, \dots, 450$. 
		}
		\label{fig:PCzzd_vs_our}
	\end{figure}
\end{center}

Figure~\ref{fig:PCzzd_vs_our} shows that over a wide range of true $m_0$ values, $\mPCNew$ and $\mZZKB$ yield comparable results both w.r.t. the point estimates and for the number of  rejections. 
In fact, $\mPCNew$ appears to be slightly more efficient than $\mZZKB$.

\begin{figure}[h]
	\centering
	\includegraphics[width=0.7\linewidth]{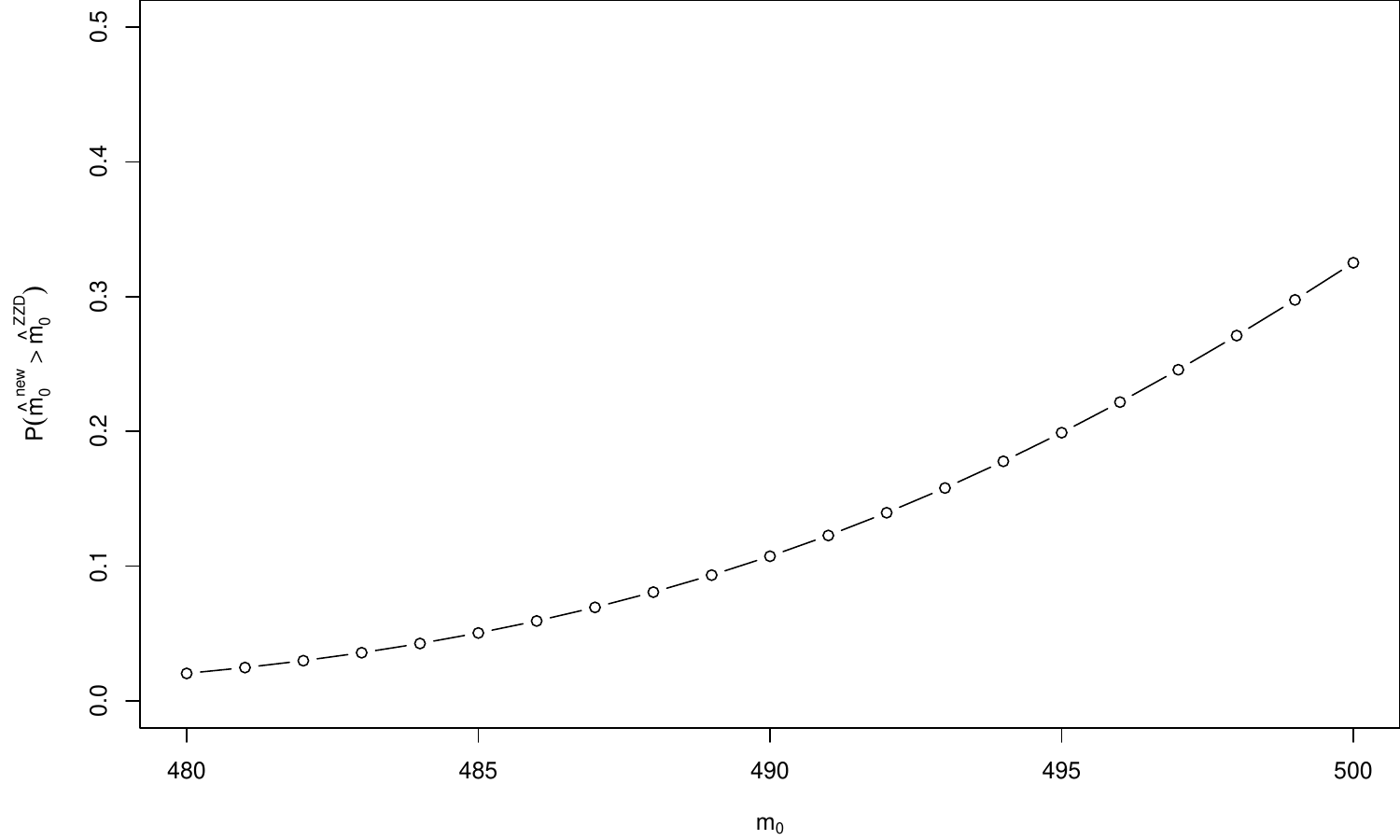}
	\caption{Approximate probabilities $\P(\mPCNew> \mZZKB)$ for various values of true $m_0$, with fixed $m=500$. }
	\label{fig:comparisonzzd}
\end{figure}

Another comparison can be obtained when we assume that the signal under the alternative is strong and that most hypotheses are nulls. 
In this case  we have $2 \sum_{i =1}^m p_i \approx 2 \sum_{i \in \nullset} p_i =:S$ so that we can use the Central Limit Theorem to quantify the probability that $\mPCNew$ is more conservative than $ \mZZKB$, i.e.
\begin{align*}
	\P (\mPCNew > \mZZKB) & = \P(S > m \cdot C(m) - 2) \approx \overline{\Phi} \left(\sqrt{\frac{3}{m_0}}\cdot (m\cdot C(m)-(m_0+2))\right).
\end{align*}
Figure \ref{fig:comparisonzzd} shows that this probability, for various values of the true $m_{0}$, is quite small and even under the complete null ($m_{0} = 500$) it is bounded by $1/3$.
By contrast, our new modification \eqref{eq:def:PC:new}  is extremely simple and, as we have shown in Section \ref{sec:discretestimators}, is easily adapted to discrete tests.

\clearpage

\end{document}